\documentclass[aps,prl,twocolumn,floatfix,superscriptaddress]{revtex4-1}
\usepackage{amsfonts,amsmath,amsthm,amssymb,amsopn,graphicx}
\usepackage{dsfont,bbm,xcolor}
\usepackage{enumitem}
\usepackage{hyperref}
\usepackage{breqn}
\usepackage{siunitx}
\usepackage{color}
\usepackage{subcaption}

\makeatletter
\let\cat@comma@active\@empty
\makeatother

\newcommand{\ii}{ {\rm i} }
\def\bra#1{\mathinner{\langle{#1}|}}
\def\ket#1{\mathinner{|{#1}\rangle}}

\newcommand{\ave}[1]{{\langle #1\rangle}}

\newcommand{\tr}{\rm{tr}}

\newcommand{\adT}{\mathcal{H}}
\def\one{\mathbbm{1}}
\def\bra#1{\mathinner{\langle{#1}|}}
\def\ket#1{\mathinner{|{#1}\rangle}}

\newcommand{\dd}{ {\rm d} }

\newcommand{\LL}{{{\cal L}}}

\newcommand{\half}{{\textstyle\frac{1}{2}}}

\def\tr{{{\rm tr}}}

\def\one{\mathbbm{1}}
\def\Re{{\,{\rm Re}\,}}
\def\Im{{\,{\rm Im}\,}}

\def\Re{{\,{\rm Re}\,}}

\newcommand{\sM}{\mathcal{M}}
\newcommand{\sL}{\mathcal{L}}

\newcommand{\Eref}[1] {Equation (\ref{#1})}

\begin{document}
\title{Dissipative Bethe Ansatz: Exact Solutions of Quantum Many-Body Dynamics Under Loss}
\author{Berislav Bu\v{c}a}
\affiliation{Clarendon Laboratory, University of Oxford, Parks Road, Oxford OX1 3PU, United Kingdom}
\author{Cameron Booker}
\affiliation{Clarendon Laboratory, University of Oxford, Parks Road, Oxford OX1 3PU, United Kingdom}
\author{Marko Medenjak}
\affiliation
{Institut de Physique Th\'eorique Philippe Meyer, \'Ecole Normale Sup\'erieure, \\ PSL University, Sorbonne Universit\'es, CNRS, 75005 Paris, France}
\author{Dieter Jaksch}
\affiliation{Clarendon Laboratory, University of Oxford, Parks Road, Oxford OX1 3PU, United Kingdom}
\begin{abstract}
We use the Bethe Ansatz technique to study dissipative systems experiencing loss. The method allows us to exactly calculate the Liouvillian spectrum. This opens the possibility of analytically calculating the dynamics of a wide range of experimentally relevant models including cold atoms subjected to one and two body losses, coupled cavity arrays with bosons escaping the cavity, and cavity quantum electrodynamics. As an example of our approach we study the relaxation properties in a boundary driven XXZ spin chain. We exactly calculate the Liouvillian gap and find different relaxation rates with a novel type of dynamical dissipative phase transition. This physically translates into the formation of a stable domain wall in the easy-axis regime despite the presence of loss. Such analytic results have previously been inaccessible for systems of this type. 
\end{abstract}		

\maketitle
\emph{\bf Introduction--} Particle loss is an important mechanism of environmental dissipation. It strongly affects the dynamics of many particle systems in the classical and the quantum regimes and has immense impact on technological applications. It is present in numerous experimental platforms including cold atoms \cite{coldexp,coldatomreview}, non-linear waveguides \cite{waveguides}, coupled cavity arrays \cite{cavityarray,Dimitris1,Dimitris2,KollathCavity}, THz cavities \cite{THzCavity,THzCavity1,KollathTHz,Frank1}, quantum wires \cite{KollathQuantumWire,Esslingerlocal1}, condensed matter systems \cite{Andrea1}, and solid-state devices \cite{solidstate}. Indeed, the primary source of dissipation in these settings is a consequence of particles escaping from the system either through one- or two-body processes \cite{coldatomreview}, or due to the coupling to an external electromagnetic field (e.g. \cite{THzCavity,solidstate,cavityarray}). 
The platforms underlie future quantum technologies, which will require efficient manipulation of many constituents. 

Understanding the behaviour of such systems is of paramount importance, and sheds light on properties that are robust to dissipation and could therefore allow for more efficient methods of information storage and the development of novel error correction mechanisms. However, due to the exponential complexity, numerical simulations of these systems are challenging. Thus, gaining a better understanding of their properties through uncovering their analytical structure is highly desirable. Thus far, exact solutions of such systems have been limited only to the stationary states of boundary driven systems \cite{Prosen1,Prosen2,Prosen3,Enej1,Enej2,Enej3,Bojan1,Bojan2,Zala1,Popkov1,Popkov2,Yuge,Lenart1,Lenart2,BucaProsen1,Nigro,ProsenReview,Diehl1,BucaProsen3} and to those with non-interacting Hamiltonians \cite{Znidaric1,Znidaric2,Znidaric3,Prosen3rd1,Prosen3rd2,Manzano1,Monthus1,Krapivsky,JuanP1,Zoller,Diehl2,Mariya1,Guo,Essler1,Lamacraft,Essler2,Katsura1,Katsura2}. Beyond this only certain approximate methods \cite{Gritsev,preprint_hydro,Zala2,Zala3}, e.g. introducing dissipation on hydrodynamical scales, are available. 

In this Letter we go beyond these results and develop an analytic approach to describing the dynamics of a wide class of fully interacting dissipative systems. Our approach opens a novel avenue for the analytical study of experimentally relevant many-body models experiencing loss, provided that the system's effective non-Hermitian Hamiltonian is integrable. In experimental settings, examples of systems treatable by our method can be found in cold atom quantum simulators subjected to single and two body losses \cite{coldexp,coldatomreview,THzCavity}, and driven-dissipative cavity arrays of bosons \cite{cavityarray}.

As an example of the power of our method we study the instructive and paradigmatic XXZ spin chain, often used to describe limiting cases of the aforementioned experimental setups \cite{coldatomreview}, which we subject to boundary spin loss. We find that our model exhibits intriguing physical phenomena. Additionally, these types of localized loss processes recently attracted a lot of theoretical and experimental interest due to their importance for understanding transport properties and as an experimentally realistic venue for preparing interesting quantum states, see e.g. \cite{Ortocat,Kuhr,Daley1,waveguides,LocalBEC,Esslingerlocal1,KollathQuantumWire,KollathQuantumWire2,Prosen1,BucaProsen1,BucaProsen,BucaProsen2,BucaProsen3,Mendoza1}. 

Using our method we first characterize the relaxation dynamics, uncovering a dynamical dissipative phase transition \cite{CiracDPT2}, by calculating the closure of the Liouvillian gap. Next, we analytically show the presence of a novel type of dynamical dissipative phase transition that corresponds to non-analiticity in many relaxation rates beyond the leading decay mode. Physically this implies a transition in the dynamics on both short and asymptotic time scales. This should be contrasted with phase transitions in the statationary state \cite{CiracDPT,Keeling1,JuanPDPT,Fazio} or the leading decay mode \cite{CiracDPT2}. In our case the stationary state is always the same and a phase transition occurs in the leading decay mode \emph{and} in other parts of the spectrum. Related to this, we show that a stable domain wall state is formed in the easy-axis regime. Interestingly, the domain wall formation occurs spontaneously if the system is initialized in the maximally polarized state. It arises as a consequence of boundary bound states that we solve for. Formation of domain walls in both integrable and non-integrable closed systems has also recently attracted considerable interest \cite{collura2018analytic,gamayun2019domain,misguich2019domain,medenjak2020domain,collura2020domain}, but is currently still analytically unsolved. 

\emph{\bf Solving lossy models--} 
We will focus on systems described by the Lindblad master equation which characterizes open quantum systems in the weak system-bath coupling limit. The dynamics of the density matrix $\rho$ is provided by the generator $\mathcal{L}$  as \cite{BPTextbook,GardinerTextbook},
\begin{eqnarray}
&&\frac{\dd}{\dd t}\rho(t) = \LL\rho(t):= \nonumber \\
&&-\ii [H,\rho(t)] +  \sum_\mu \left(2L_\mu \rho(t) L_\mu^\dagger - \{L_\mu^\dagger L_\mu,\rho(t)\}\right),
\label{eq:lindeq}
\end{eqnarray}
where $H$ is the system Hamiltonian and $L_\mu$ are the Lindblad jump operators modeling the influence of the environment on the system. The time evolution of an observable $O$ can be computed by diagonalizing the generator $\LL$, $\ave{O(t)}=\tr( O \rho(t) )=\sum_\mu e^{\lambda_\mu t} \tr(\tilde{\rho}^\dagger_\mu \rho(0)) \tr(\rho_\mu O)$, where $\lambda_\mu$ are the eigenvalues and $\rho_\mu, \tilde{\rho}_\mu$ the right and left eigenvectors.

The general setup that we consider comprises an integrable Hamiltonian $H$ with a conservation law $M$ and Lindblad operators $L_\mu$ that change the eigenvalue of $M$ by well-defined amounts $m_\mu>0$,  $[M,L_\mu]= -m_\mu L_\mu$, inducing the loss of the quantity $ M $ in the system. For instance $M$ can be the total particle number and $L_\mu$ particle annihilation operators. In the following we will discuss the integrability requirements for applying our technique. The Liouvillian superoperator can be represented on the vector space with doubled degrees of freedom by the channel-state transformation  $\ket{\psi}\bra{\phi} \to \ket{\psi}\otimes\ket{\phi}$, yielding
\begin{dmath} \label{eq:doublespaceliouvillian}
\LL=-i(H\otimes\one-\one\otimes H^T)+ +\sum_\mu \left(2 L_\mu\otimes L_\mu^*-L_\mu^\dagger L_\mu\otimes \one-\one \otimes (L_\mu^\dagger L_\mu)^T\right).
\end{dmath}

We will show that in order to obtain eigenvalues of the Liouvillian it suffices to obtain eigenvalues $ E_j $ of the non-hermitian Hamiltonian $\tilde{H}\equiv-\ii H- \sum_\mu L_\mu^\dagger  L_\mu$ \cite{Torres1,Englert}. Since $ [\tilde{H},M]=0 $, we can assume that the eigenvectors $\ket{\psi_j} $ of $\tilde{H}$  are also eigenvectors of $ M $. The generator \eqref{eq:doublespaceliouvillian} can now be decomposed into two parts 
\begin{equation}\label{eq3}
 \LL=\mathcal{H}+\mathcal{D},
\end{equation}
  with $\mathcal{H}\equiv\tilde{H}\otimes\one+\one\otimes\tilde{H}^*$ and $\mathcal{D}=2\sum L_\mu\otimes L_\mu^*$. Since $ \adT $ is a sum of two operators acting on the factors in a tensor product independently, its eigenvalues read
  $
  	\adT \ket{\psi_i} \otimes \overline{\ket{\psi_j}} = (E_i +E_j^*) \ket{\psi_i}\otimes \overline{\ket{\psi_j}}.$
Let us now order the eigenvectors $ \ket{\psi_i} \otimes \overline{\ket{\psi_j}} $ of $\adT$ by the corresponding eigenvalues $ m_{i,j} $ of $ \mathcal{M}\equiv M\otimes\one+\one\otimes M $. Due to the purely lossy dynamics the nondiagonal matrix elements of $ \mathcal{D} $ lie strictly above the diagonal. This immediately implies that $ \LL $ takes the upper triangular form in the basis $ \ket{\psi_i}\otimes \overline{\ket{\psi_j}} $ and that the eigenvalues of the Liouvillian $\lambda_{i,j} = E_i +E_j^*$ coincide with those of $\adT$.
Thus, provided that the non-hermitian Hamiltonian $\tilde{H}=-\ii H- \sum_\mu L_\mu^\dagger L_\mu$ is exactly solvable, we have found the full spectrum of the Liouvillian.
Nevertheless, the structure of Liouvillian eigenvectors corresponding to the eigenvalue $\lambda_{i,j}$ is more complicated and includes the states $ \ket{\psi_k}\otimes \overline{\ket{\psi_l}} $, with $m_{k,l}\leq m_{i,j}$ \cite{SM}.

Note that the dynamics describing pure gain can be treated on the same footing.

Here we focus on using Bethe Ansatz techniques \cite{Bethe1,Bethe2} for solving $\tilde{H}$, which are applicable to a wide range of models. In many physically relevant situations the dissipative contribution, $ \sum_\mu L_\mu^\dagger L_\mu $, modifying the system's integrable Hamiltonian, $H$, will leave $\tilde{H}$ integrable. For instance, single particle bulk loss throughout the system in \emph{any} integrable model with particle number conservation, the dissipative contribution corresponds simply to the particle number operator, which clearly implies integrability of $ \tilde{H} $. For two-level systems with conserved magnetization the $L_\mu$ would correspond to on-site spin lowering operators. This is known to be a primary dissipative loss mechanism in numerous experimental setups such as optical lattices (due to interactions with the background vacuum), wave guides, solid state contacts, and coupled cavity arrays \cite{coldatomreview,waveguides,cavityarray,solidstate}.
Other examples of dissipative mechanisms that preserve integrability of $\tilde{H}$ are provided by nearest-neighbour dissipation \cite{Chris,Chris2} and two-body loss processes \cite{coldatomreview,Ueda} (for details see \cite{SM}). 

It is instructive to contrast this situation with cases where the full Liouvillian can be mapped to a non-Hermitian integrable Hamiltonian \cite{Essler1,Essler2,Lamacraft,Katsura1,Katsura2} (where the physical system's Hamiltonian is quadratic). In our case the system's Hamiltonian is interacting and the full Liouvillian does not correspond to some non-Hermitian integrable Hamiltonian. Rather here it is only $\tilde{H}$ (and hence $\adT$ in Eq.~\eqref{eq3}) that is integrable. For conciseness and in order to demonstrate the utility of our method we will now focus on the example of the Heisenberg XXZ chain in the presence of a spin sink at a single boundary. 

\emph{\bf Boundary driven XXZ spin chain dynamics--} 
The Heisenberg Hamiltonian reads
\begin{align}
\label{Heisenberg}
& H_{XXZ} =\sum_{j=1}^{N-1} \sigma^x_j \sigma^x_{j+1}+ \sigma^y_j \sigma^y_{j+1}+\Delta \sigma^z_j \sigma^z_{j+1},
\end{align}
where the Pauli spin$ -\half $ operators are $\sigma^{x,y,z}$, $\Delta $ is the anisotropy, and $N$ is the number of spin sites. We study the setup with an arbitrary loss rate on the first site $L_1=2\sqrt{ \Gamma} \sigma^-_1$. The corresponding non-Hermitian Hamiltonian reads
\begin{equation}
\label{nonhermham}
\tilde{H}=-\ii H_{XXZ} -4\Gamma \sigma^z_1 +\rm{const},
\end{equation}
i.e. $\ii \tilde{H}$ describes an XXZ spin chain Hamiltonian in the presence of an imaginary magnetic field at the boundary. The Hamiltonian has a $U(1)$ symmetry $M=\sum_j \sigma^z_j$ and $[M,L_1]=-L_1$, i.e. $m_1=1$. 

In contrast to boundary driven spin chains \cite{Prosen1,Prosen2,Prosen3,Znidaric1,Bojan3,Clark1,Clark2,Clark3,Clark4} the stationary state, $\LL\rho_{\infty} =0$, is not of interest in our system since it is a trivial vacuum state.  However, we obtain the full spectrum of the non-Hermitian Hamiltonian  Eq.~\eqref{nonhermham}, and therefore of the Liouvillian by the Bethe ansatz. The Bethe equations were obtained using Sklyanin's reflection algebra \cite{Sklyanin}, and equivalently the coordinate Bethe ansatz \cite{Bethe2,Ragoucy1,Ragoucy2} (with imaginary boundary magnetic field $4 i \Gamma \sigma^z_1$).

The complex energies of $ \tilde{H} $ corresponding to $ m $ magnons read
\begin{equation} \label{eq:bethe_pseudo_energies}
    E(\{k_j\}) = -\ii(N-1) -4\ii \sum_{j=1}^{m} (\cos(k_j)-\Delta),
\end{equation}
where the momenta of the magnons, $\{k_j\}$, are obtained by solving the Bethe equations
\begin{equation} \label{eq:bethe_equations}
     \frac{e^{2iNk_j}(\Delta -e^{ik_j})(e^{ik_j} +2i\Gamma -\Delta)}{(e^{ik_j} \Delta -1)(1+e^{ik_j}(2i\Gamma -\Delta))} = \prod_{l\ne j}^m S(e^{ik_j}, e^{ik_l}).
\end{equation}
The scattering matrix of two magnons takes the form
\begin{equation}
    S(a,b) = \frac{(a-2\Delta ab+b)(1-2\Delta a +ab)}{(a-2\Delta +b)(1-2\Delta b +ab)}.
\end{equation}
 From the triangular form of the Liouvillian, which couples different magnetization sectors, we can express its eigenstates in terms of Bethe states of $\tilde{H}$ by simplified Gaussian elimination (for details see \cite{SM}). For later convenience we introduce a pair of labels $(m_L,m_R)$ refering to the eigenstates of $\LL$ comprised of tensor product of two Bethe states with $m_L$ and $m_R$ magnons as well as tensor products of Bethe states with less $q_L<m_L$ and $q_R<m_R$ magnons  (see SI \cite{SM}). In what follows we will utilise our results to address two physically interesting questions. 

\begin{figure}[!]
\begin{center}
\vspace{0mm}
    \includegraphics[width = \columnwidth]{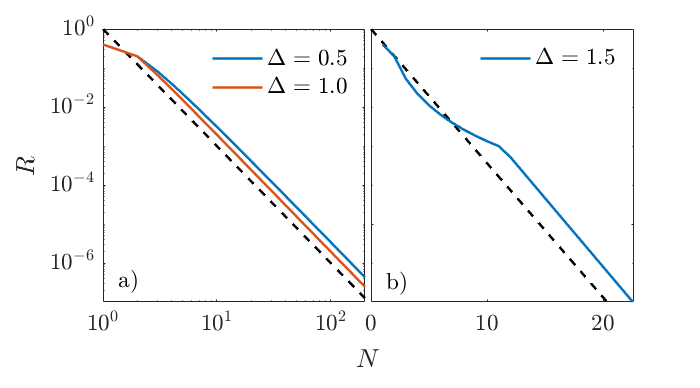}
\end{center}
	\vspace{-8mm}
    \caption{Scaling of the Liouvillian gap, $R$, with system size for $\Gamma = 0.1$. a) Power law closing of the gap for $\Delta\leq 1$ with $R\propto 1/N^3$ (dashed line) shown for comparison. b) Exponential closing of the gap for $\Delta>1$ with $R\propto e^{-\alpha N}$ (dashed line) and $\alpha \approx 0.784$ shown for comparison. }
    \label{fig:liouvilliangap}
    
\end{figure}

\emph{\bf Eigenvalue Structure and the Liouvillian gap---}
The first problem that we consider is the Liouvillian gap, $R$, of $\LL$. It corresponds to the maximum real part of the eigenvalues different from 0, which is the inverse relaxation time of the longest-lived eigenmodes. We plot it in Fig. \ref{fig:liouvilliangap} for different values of the anisotropy parameter $\Delta$. 
The scaling of the gap with the system size $N$ is one of the primary features of open quantum systems, governing the late time dynamics. 

\begin{figure*}[t]
	\begin{subfigure}{0.32\textwidth}
		\centering
		\includegraphics[width=\textwidth]{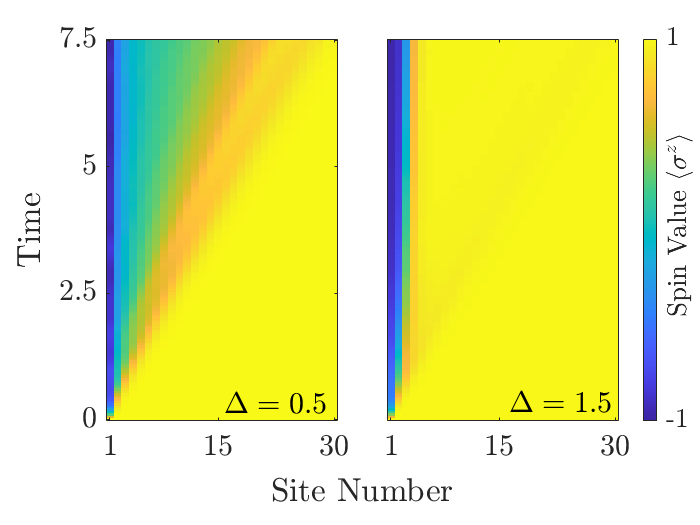}
		\caption{}
		\label{fig:domainwall&decay}
	\end{subfigure}
	\begin{subfigure}{0.32\textwidth}
	    \centering
	    \includegraphics[width=\textwidth]{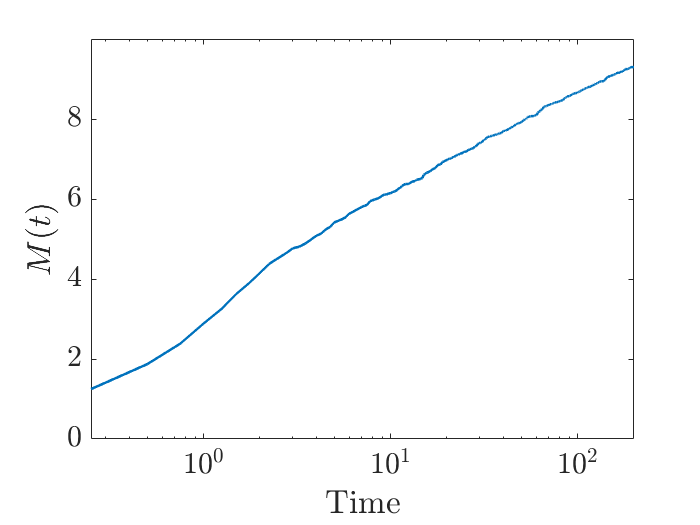}
	    \caption{}
	    \label{fig:magnetisationdecay}
	\end{subfigure}
		\begin{subfigure}{0.32\textwidth}
		\centering
		\includegraphics[width=\textwidth]{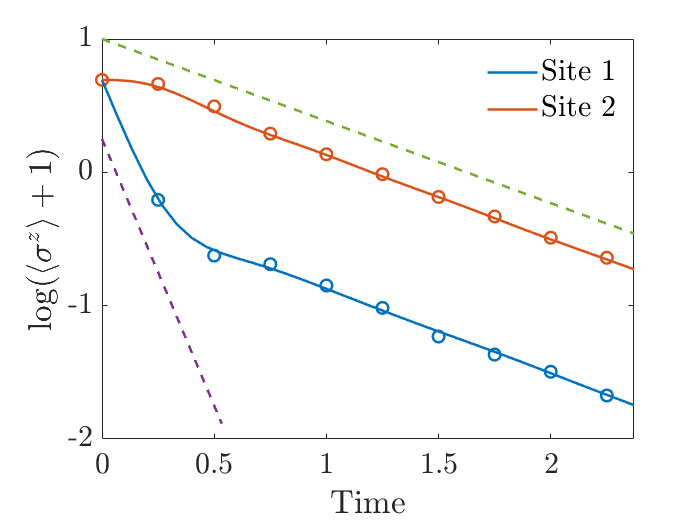}
		\caption{}
		\label{fig:domainwall_spinloss}
	\end{subfigure}
    \caption{Domain wall formation and spin loss. a) Magnetization along the chain as a function of time for an initially fully polarized state with $\Gamma = 0.5$. For $\Delta=0.5$ the initial state rapidly decays to the vacuum state $\rho_\infty$, while for $\Delta=1.5$ the system forms a meta-stable domain wall which decays exponentially slowly.
    b) The total spin lost, $M(t)$, by a 30 site system with $\Gamma = 0.5$. This is logarithmic in time as a result of the domain wall preventing spin from leaking out. c) The spin decay on the first and second sites in 10 (solid) and 30 (circles) site systems with $\Delta = 1.5$, $\Gamma = 0.5$. We also plot the infinite chain 0 and 1 top-magnon decay rates in purple and green respectively for comparison.}
	\label{fig:tDMRGplots}
\end{figure*}
We observe that the gap for $\Delta\leq1$ corresponds to the eigenstates of $\LL$ with $m_L+m_R=1$.
By examining the single magnon $m=1$ solutions of \eqref{eq:bethe_equations} on top of the steady state we find that, depending on the value of $\Delta$, the gap closes at different rates. In particular, we demonstrate that for $\Delta \le 1$ the longest lived excitations correspond to the solutions with $\lim_{N \to \infty}\Im(k_j)=0$. Rewriting Eq.~\eqref{eq:bethe_equations} as,
$
2 i k=\frac{\Omega(e^{\ii k}) }{N}-\frac{2 \ii \pi  I_1}{N},
$
with $\Omega(a)= \log \left(\frac{(a \Delta -1) (-1+a (\Delta -2 \ii \Gamma ))}{(a-\Delta ) (a+2 \ii \Gamma -\Delta )}\right)$, we find that for $\Delta \le 1$ the real part of the eigenvalues with the smallest real part scale as, \cite{SM},
\begin{equation}
R=-\frac{1}{N^3}\frac{8 \pi ^2 \Gamma }{4 \Gamma ^2+(\Delta +1)^2}+{\cal O}(\frac{1}{N^4}). \label{gapeq}
\end{equation}
This matches the scaling of the gap for free fermions \cite{Prosen3rd1}. In the $N\to\infty$ limit these solutions are straightforwardly generalized to $m$-magnons \cite{SM}. However, for $\Delta>1$ the leading decay rate is not in this class of solutions. Instead, we find exponentially long relaxation times consistent with a gap that closes exponentially fast (see Fig. \ref{fig:liouvilliangap} and \cite{SM}).

\emph{\bf Boundary bound states and domain wall formation in the easy-axis regime---}
In the second setup we consider the case where the system is initialized in a highly excited, i.e. maximally polarized (all spins-up) state. In this case, due to the structure of $\LL$, we need only consider eigenstates with $m_L = m_R = m$ (see \cite{SM}). In order to study the dynamics we now focus on the most stable (maximum real part) eigenvalues in the $m$ \emph{top-magnon} sector, corresponding to spin-down excitations on top of the all spins-up state. The Bethe equations for top-magnons can be obtained from Eq.~\eqref{eq:bethe_equations} by replacing $\Gamma\to-\Gamma$ in the sector with $m$ magnons.

Focusing on the easy-axis, $\Delta>1$, regime, we show that in the $m$ top-magnon sector states with $\lim_{N \to \infty}\Im(k_j)>0$, which are localized at the boundary appear and are the most stable (see SI \cite{SM}).
For these \emph{bound} states the $m$ top-magnon Bethe equations can be easily solved in the $N \to \infty$ limit, since $e^{\ii k_j N} \to 0$ (see SI \cite{SM}). A recursive solution of
\begin{eqnarray*}
&&\exp(-\ii k_j)+\exp(i k_{j-1})=2 \Delta,\\ &&\exp(-\ii k_1)=\Delta+2\ii \Gamma,
\end{eqnarray*}
gives an appealingly simple result for the leading Liouvillian eigenvalues in the $m$ top-magnon sector,
\begin{equation}
\lambda_m=-2 \ii(\exp(\ii k_m)-\exp(-\ii k^*_m)). \label{eigenvalue}
\end{equation}
Physically this means that the first top-magnon with momentum $k_1$ is localized near the loss site, while the $j$th top-magnon is recursively bound to the $(j-1)$st. Importantly, we can show that as the number of top-magnons is increased they become exponentially stabilized, i.e. $\lim_{m\to\infty}\Re{\lambda_m}=0$ (see SI \cite{SM}). In turn, this implies that exponentially large times (in $m$) are needed for the loss site to dissipate the state with $m$ top-magnons.

The existence of these boundary bound states has intriguing physical consequences. It results in domain wall formation if the system is initialized in the maximally polarized state. Naively one might think that such a state is the most unstable, however tDMRG simulations, as shown in Fig.~\ref{fig:domainwall&decay}, in the $\Delta>1$ regime reveal that the total spin leaking out of the system increases only \emph{logarithmically} with time (see Fig.~\ref{fig:magnetisationdecay}). This can be understood as a consequence of exponential stability of the boundary bound states. Namely, that exponentially long times (in $m$) are required for the loss site to remove all the states with $m$ down-turned spins. Moreover, in Fig.~\ref{fig:domainwall_spinloss} we show that the dynamics of magnetization close to the spin loss site is well described by the decay rates of boundary top-magnons. On the other hand the decay of the maximally polarized state in the $\Delta<1$ regime is very rapid (Fig.~\ref{fig:domainwall&decay}), and the total loss of magnetization increases linearly with time. There have recently been a number of studies addressing the dynamics of domain walls in integrable \cite{collura2018analytic,gamayun2019domain,misguich2019domain,collura2020domain} and nonintegrable \cite{medenjak2020domain} systems without dissipation. While the ballistic expansion in the $ \Delta<1 $ regime is well understood, the domain wall freezing was analytically unresolved.

\begin{figure}[t]
    \centering
    \includegraphics[width=\columnwidth]{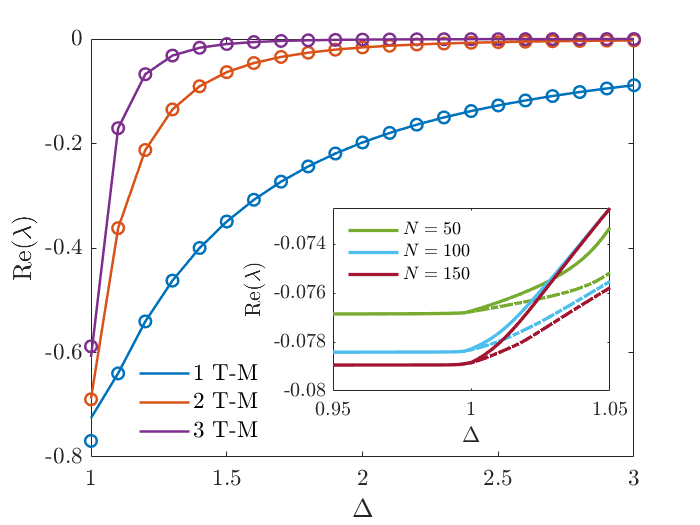}
    \caption{The eigenvalues with maximum real part of $\LL$ when $N\rightarrow\infty$ (circles) from \Eref{eigenvalue} for $\Delta>1$ compared with a 50 site system (line) for 1, 2 and 3 top-magnon sectors at $\Gamma = 0.01$. The inset shows the largest (solid line) and second largest (dashed line) real part of the eigenvalues of the Liouvillian in the one top-magnon sector for different $N$ at $\Gamma = 0.01$. We see a cusp forming with increasing $N$ close to $\Delta=1$ indicating the dynamical dissipative phase transition in the large $N$ limit.}
    \label{fig:phase_transition}
\end{figure}

The existence of boundary bound states also has profound consequences on the spectral properties of $\LL$. It results in a dissipative phase transition (shown in Fig.~\ref{fig:phase_transition}). In contrast to standard dissipative phase transitions \cite{CiracDPT}, the stationary state remains the same (all spins-down). The phase transition rather happens in the relaxation spectrum of $\LL$ at different values of $\Delta$ and $\Gamma$ depending on the top-magnon number $m$, and converges to $\Delta=1$ in the limit $m\to\infty$. This is similar to \emph{dynamical} dissipative phase transitions \cite{CiracDPT2}, but the discontinuous eigenvalues that are relevant for the dynamics are not only the Liouvillian gap. This is reflected in the fact that already the short time dynamics for the easy-plane and easy-axis regimes are qualitatively different (see Fig.~\ref{fig:tDMRGplots}) The discontinuity is shown in Fig.~\ref{fig:phase_transition} where we can see non-analyticity in eigenvalues in three different top-magnon sectors demonstrating that this phase transitions happens in all sectors. The non-analyticity shown corresponds to the non-existence of boundary bound state solutions for $\Delta<1$.  More specifically, we prove the existence of $\{k_j\}$ such that $\lim_{\Delta \to 1}\frac{d k_j}{d \Delta} \to \infty $ for large $N$ and small $\Gamma$ (see SI \cite{SM}), which implies the divergence in the corresponding eigenvalues.

\emph{\bf Conclusion---} We have devoloped a framework for diagonalizing quantum Liouvillians with integrable system Hamiltonians and dissipative loss. We demonstrate the utility of our method in an example of the Heisenberg XXZ spin chain with boundary loss. The method allows us to directly identify phase transitions in the Liouvillian spectrum and calculate the Liouvillian gap. This led us to observe two intriguing physical phenomena, namely domain wall formation, and a dissipative phase transition, which we link to the existence of boundary bound top-magnons. Such remarkable phenomena could occur in other models with localized loss, e.g. 1D Hubbard and interacting bosons in 1D \cite{coldatomreview}, which can be studied analytically with our method.

A number of questions remain open. The first natural extension of our results is directly calculating the full eigenstates of the quantum Liouvillian. We also envisage using the thermodynamic Bethe ansatz \cite{van2016introduction,YangYang} to explore the decay of states with a finite density of excitations, and the connection with boundary states \cite{Jacopo2} and strong edge modes \cite{Fendley1} in closed systems. Additionally, the Liouvillian spectrum exhibits a multi-band structure at sufficiently large $\Delta$ (see Fig. \ref{fig:fullspect} in \cite{SM}), which remains to be explained.

More generally our method can be applied to a number of systems that are quantitatively very different from the example we studied. Such systems include, for example, arrays of two-level systems with nearest-neighbor dissipation induced by external drive \cite{Chris}, and integrable systems exhibiting the loss of particles at each site \cite{coldatomreview}. Here the interest is two-fold. On one hand, judging by our example, such systems hide a plethora of interesting physical phenomena, which are yet to be uncovered. On the other hand they describe realistic experimental setups and therefore provide an indispensable tool for understanding future experiments.

\emph{Note Added:} While nearing the completion of this manuscript a related preprint appeared \cite{Ueda} studying exact solutions in the Hubbard model with two-body loss. 

\emph{Acknowledgments---}  We thank F. Essler, E. Ilievski, C. Parmee, T. Prosen, J. Tindall, F. Tonielli, and A. Ziolkowska for useful discussions. We are in particularly indebted to T. Prosen for suggesting the original problem that led to this study. BB, CB and DJ acknowledge funding from EPSRC programme grant EP/P009565/1, EPSRC National Quantum Technology Hub in Networked Quantum   Information Technology (EP/M013243/1), and the European Research Council under the European Union's Seventh Framework Programme (FP7/2007-2013)/ERC Grant Agreement no. 319286, Q-MAC. 

\bibliographystyle{apsrev4-1}
\bibliography{main}

\begin{thebibliography}{97}%
\makeatletter
\providecommand \@ifxundefined [1]{%
 \@ifx{#1\undefined}
}%
\providecommand \@ifnum [1]{%
 \ifnum #1\expandafter \@firstoftwo
 \else \expandafter \@secondoftwo
 \fi
}%
\providecommand \@ifx [1]{%
 \ifx #1\expandafter \@firstoftwo
 \else \expandafter \@secondoftwo
 \fi
}%
\providecommand \natexlab [1]{#1}%
\providecommand \enquote  [1]{``#1''}%
\providecommand \bibnamefont  [1]{#1}%
\providecommand \bibfnamefont [1]{#1}%
\providecommand \citenamefont [1]{#1}%
\providecommand \href@noop [0]{\@secondoftwo}%
\providecommand \href [0]{\begingroup \@sanitize@url \@href}%
\providecommand \@href[1]{\@@startlink{#1}\@@href}%
\providecommand \@@href[1]{\endgroup#1\@@endlink}%
\providecommand \@sanitize@url [0]{\catcode `\\12\catcode `\$12\catcode
  `\&12\catcode `\#12\catcode `\^12\catcode `\_12\catcode `\%12\relax}%
\providecommand \@@startlink[1]{}%
\providecommand \@@endlink[0]{}%
\providecommand \url  [0]{\begingroup\@sanitize@url \@url }%
\providecommand \@url [1]{\endgroup\@href {#1}{\urlprefix }}%
\providecommand \urlprefix  [0]{URL }%
\providecommand \Eprint [0]{\href }%
\providecommand \doibase [0]{http://dx.doi.org/}%
\providecommand \selectlanguage [0]{\@gobble}%
\providecommand \bibinfo  [0]{\@secondoftwo}%
\providecommand \bibfield  [0]{\@secondoftwo}%
\providecommand \translation [1]{[#1]}%
\providecommand \BibitemOpen [0]{}%
\providecommand \bibitemStop [0]{}%
\providecommand \bibitemNoStop [0]{.\EOS\space}%
\providecommand \EOS [0]{\spacefactor3000\relax}%
\providecommand \BibitemShut  [1]{\csname bibitem#1\endcsname}%
\let\auto@bib@innerbib\@empty
\bibitem [{\citenamefont {Gross}\ and\ \citenamefont {Bloch}(2017)}]{coldexp}%
  \BibitemOpen
  \bibfield  {author} {\bibinfo {author} {\bibfnamefont {C.}~\bibnamefont
  {Gross}}\ and\ \bibinfo {author} {\bibfnamefont {I.}~\bibnamefont {Bloch}},\
  }\href {\doibase 10.1126/science.aal3837} {\bibfield  {journal} {\bibinfo
  {journal} {Science}\ }\textbf {\bibinfo {volume} {357}},\ \bibinfo {pages}
  {995} (\bibinfo {year} {2017})}\BibitemShut {NoStop}%
\bibitem [{\citenamefont {Lewenstein}\ \emph {et~al.}(2007)\citenamefont
  {Lewenstein}, \citenamefont {Sanpera}, \citenamefont {Ahufinger},
  \citenamefont {Damski}, \citenamefont {Sen(De)},\ and\ \citenamefont
  {Sen}}]{coldatomreview}%
  \BibitemOpen
  \bibfield  {author} {\bibinfo {author} {\bibfnamefont {M.}~\bibnamefont
  {Lewenstein}}, \bibinfo {author} {\bibfnamefont {A.}~\bibnamefont {Sanpera}},
  \bibinfo {author} {\bibfnamefont {V.}~\bibnamefont {Ahufinger}}, \bibinfo
  {author} {\bibfnamefont {B.}~\bibnamefont {Damski}}, \bibinfo {author}
  {\bibfnamefont {A.}~\bibnamefont {Sen(De)}}, \ and\ \bibinfo {author}
  {\bibfnamefont {U.}~\bibnamefont {Sen}},\ }\href {\doibase
  10.1080/00018730701223200} {\ \textbf {\bibinfo {volume} {56}},\ \bibinfo
  {pages} {243} (\bibinfo {year} {2007})}\BibitemShut {NoStop}%
\bibitem [{\citenamefont {Zezyulin}\ \emph {et~al.}(2012)\citenamefont
  {Zezyulin}, \citenamefont {Konotop}, \citenamefont {Barontini},\ and\
  \citenamefont {Ott}}]{waveguides}%
  \BibitemOpen
  \bibfield  {author} {\bibinfo {author} {\bibfnamefont {D.~A.}\ \bibnamefont
  {Zezyulin}}, \bibinfo {author} {\bibfnamefont {V.~V.}\ \bibnamefont
  {Konotop}}, \bibinfo {author} {\bibfnamefont {G.}~\bibnamefont {Barontini}},
  \ and\ \bibinfo {author} {\bibfnamefont {H.}~\bibnamefont {Ott}},\ }\href
  {\doibase 10.1103/PhysRevLett.109.020405} {\bibfield  {journal} {\bibinfo
  {journal} {Phys. Rev. Lett.}\ }\textbf {\bibinfo {volume} {109}},\ \bibinfo
  {pages} {020405} (\bibinfo {year} {2012})}\BibitemShut {NoStop}%
\bibitem [{\citenamefont {Fitzpatrick}\ \emph {et~al.}(2017)\citenamefont
  {Fitzpatrick}, \citenamefont {Sundaresan}, \citenamefont {Li}, \citenamefont
  {Koch},\ and\ \citenamefont {Houck}}]{cavityarray}%
  \BibitemOpen
  \bibfield  {author} {\bibinfo {author} {\bibfnamefont {M.}~\bibnamefont
  {Fitzpatrick}}, \bibinfo {author} {\bibfnamefont {N.~M.}\ \bibnamefont
  {Sundaresan}}, \bibinfo {author} {\bibfnamefont {A.~C.~Y.}\ \bibnamefont
  {Li}}, \bibinfo {author} {\bibfnamefont {J.}~\bibnamefont {Koch}}, \ and\
  \bibinfo {author} {\bibfnamefont {A.~A.}\ \bibnamefont {Houck}},\ }\href
  {\doibase 10.1103/PhysRevX.7.011016} {\bibfield  {journal} {\bibinfo
  {journal} {Phys. Rev. X}\ }\textbf {\bibinfo {volume} {7}},\ \bibinfo {pages}
  {011016} (\bibinfo {year} {2017})}\BibitemShut {NoStop}%
\bibitem [{\citenamefont {Tangpanitanon}\ \emph {et~al.}(2016)\citenamefont
  {Tangpanitanon}, \citenamefont {Bastidas}, \citenamefont {Al-Assam},
  \citenamefont {Roushan}, \citenamefont {Jaksch},\ and\ \citenamefont
  {Angelakis}}]{Dimitris1}%
  \BibitemOpen
  \bibfield  {author} {\bibinfo {author} {\bibfnamefont {J.}~\bibnamefont
  {Tangpanitanon}}, \bibinfo {author} {\bibfnamefont {V.~M.}\ \bibnamefont
  {Bastidas}}, \bibinfo {author} {\bibfnamefont {S.}~\bibnamefont {Al-Assam}},
  \bibinfo {author} {\bibfnamefont {P.}~\bibnamefont {Roushan}}, \bibinfo
  {author} {\bibfnamefont {D.}~\bibnamefont {Jaksch}}, \ and\ \bibinfo {author}
  {\bibfnamefont {D.~G.}\ \bibnamefont {Angelakis}},\ }\href {\doibase
  10.1103/PhysRevLett.117.213603} {\bibfield  {journal} {\bibinfo  {journal}
  {Phys. Rev. Lett.}\ }\textbf {\bibinfo {volume} {117}},\ \bibinfo {pages}
  {213603} (\bibinfo {year} {2016})}\BibitemShut {NoStop}%
\bibitem [{\citenamefont {Tangpanitanon}\ \emph {et~al.}(2019)\citenamefont
  {Tangpanitanon}, \citenamefont {Clark}, \citenamefont {Bastidas},
  \citenamefont {Fazio}, \citenamefont {Jaksch},\ and\ \citenamefont
  {Angelakis}}]{Dimitris2}%
  \BibitemOpen
  \bibfield  {author} {\bibinfo {author} {\bibfnamefont {J.}~\bibnamefont
  {Tangpanitanon}}, \bibinfo {author} {\bibfnamefont {S.~R.}\ \bibnamefont
  {Clark}}, \bibinfo {author} {\bibfnamefont {V.~M.}\ \bibnamefont {Bastidas}},
  \bibinfo {author} {\bibfnamefont {R.}~\bibnamefont {Fazio}}, \bibinfo
  {author} {\bibfnamefont {D.}~\bibnamefont {Jaksch}}, \ and\ \bibinfo {author}
  {\bibfnamefont {D.~G.}\ \bibnamefont {Angelakis}},\ }\href {\doibase
  10.1103/PhysRevA.99.043808} {\bibfield  {journal} {\bibinfo  {journal} {Phys.
  Rev. A}\ }\textbf {\bibinfo {volume} {99}},\ \bibinfo {pages} {043808}
  (\bibinfo {year} {2019})}\BibitemShut {NoStop}%
\bibitem [{\citenamefont {Wolff}\ \emph {et~al.}(2016)\citenamefont {Wolff},
  \citenamefont {Sheikhan},\ and\ \citenamefont {Kollath}}]{KollathCavity}%
  \BibitemOpen
  \bibfield  {author} {\bibinfo {author} {\bibfnamefont {S.}~\bibnamefont
  {Wolff}}, \bibinfo {author} {\bibfnamefont {A.}~\bibnamefont {Sheikhan}}, \
  and\ \bibinfo {author} {\bibfnamefont {C.}~\bibnamefont {Kollath}},\ }\href
  {\doibase 10.1103/PhysRevA.94.043609} {\bibfield  {journal} {\bibinfo
  {journal} {Phys. Rev. A}\ }\textbf {\bibinfo {volume} {94}},\ \bibinfo
  {pages} {043609} (\bibinfo {year} {2016})}\BibitemShut {NoStop}%
\bibitem [{\citenamefont {Zhang}\ \emph {et~al.}(2016)\citenamefont {Zhang},
  \citenamefont {Lou}, \citenamefont {Li}, \citenamefont {Reno}, \citenamefont
  {Pan}, \citenamefont {Watson}, \citenamefont {Manfra},\ and\ \citenamefont
  {Kono}}]{THzCavity}%
  \BibitemOpen
  \bibfield  {author} {\bibinfo {author} {\bibfnamefont {Q.}~\bibnamefont
  {Zhang}}, \bibinfo {author} {\bibfnamefont {M.}~\bibnamefont {Lou}}, \bibinfo
  {author} {\bibfnamefont {X.}~\bibnamefont {Li}}, \bibinfo {author}
  {\bibfnamefont {J.~L.}\ \bibnamefont {Reno}}, \bibinfo {author}
  {\bibfnamefont {W.}~\bibnamefont {Pan}}, \bibinfo {author} {\bibfnamefont
  {J.~D.}\ \bibnamefont {Watson}}, \bibinfo {author} {\bibfnamefont {M.~J.}\
  \bibnamefont {Manfra}}, \ and\ \bibinfo {author} {\bibfnamefont
  {J.}~\bibnamefont {Kono}},\ }\href {\doibase 10.1038/nphys3850} {\bibfield
  {journal} {\bibinfo  {journal} {Nature Physics}\ }\textbf {\bibinfo {volume}
  {12}},\ \bibinfo {pages} {1005} (\bibinfo {year} {2016})}\BibitemShut
  {NoStop}%
\bibitem [{\citenamefont {Scalari}\ \emph {et~al.}(2012)\citenamefont
  {Scalari}, \citenamefont {Maissen}, \citenamefont {Tur{\v c}inkov{\'a}},
  \citenamefont {Hagenm{\"u}ller}, \citenamefont {De~Liberato}, \citenamefont
  {Ciuti}, \citenamefont {Reichl}, \citenamefont {Schuh}, \citenamefont
  {Wegscheider}, \citenamefont {Beck},\ and\ \citenamefont
  {Faist}}]{THzCavity1}%
  \BibitemOpen
  \bibfield  {author} {\bibinfo {author} {\bibfnamefont {G.}~\bibnamefont
  {Scalari}}, \bibinfo {author} {\bibfnamefont {C.}~\bibnamefont {Maissen}},
  \bibinfo {author} {\bibfnamefont {D.}~\bibnamefont {Tur{\v c}inkov{\'a}}},
  \bibinfo {author} {\bibfnamefont {D.}~\bibnamefont {Hagenm{\"u}ller}},
  \bibinfo {author} {\bibfnamefont {S.}~\bibnamefont {De~Liberato}}, \bibinfo
  {author} {\bibfnamefont {C.}~\bibnamefont {Ciuti}}, \bibinfo {author}
  {\bibfnamefont {C.}~\bibnamefont {Reichl}}, \bibinfo {author} {\bibfnamefont
  {D.}~\bibnamefont {Schuh}}, \bibinfo {author} {\bibfnamefont
  {W.}~\bibnamefont {Wegscheider}}, \bibinfo {author} {\bibfnamefont
  {M.}~\bibnamefont {Beck}}, \ and\ \bibinfo {author} {\bibfnamefont
  {J.}~\bibnamefont {Faist}},\ }\href {\doibase 10.1126/science.1216022}
  {\bibfield  {journal} {\bibinfo  {journal} {Science}\ }\textbf {\bibinfo
  {volume} {335}},\ \bibinfo {pages} {1323} (\bibinfo {year}
  {2012})}\BibitemShut {NoStop}%
\bibitem [{\citenamefont {Halati}\ \emph {et~al.}(2019)\citenamefont {Halati},
  \citenamefont {Sheikhan}, \citenamefont {Ritsch},\ and\ \citenamefont
  {Kollath}}]{KollathTHz}%
  \BibitemOpen
  \bibfield  {author} {\bibinfo {author} {\bibfnamefont {C.-M.}\ \bibnamefont
  {Halati}}, \bibinfo {author} {\bibfnamefont {A.}~\bibnamefont {Sheikhan}},
  \bibinfo {author} {\bibfnamefont {H.}~\bibnamefont {Ritsch}}, \ and\ \bibinfo
  {author} {\bibfnamefont {C.}~\bibnamefont {Kollath}},\ }\href@noop {}
  {\enquote {\bibinfo {title} {Dissipative generation of highly entangled
  states of light and matter},}\ } (\bibinfo {year} {2019}),\ \Eprint
  {http://arxiv.org/abs/1909.07335} {arXiv:1909.07335 [cond-mat.quant-gas]}
  \BibitemShut {NoStop}%
\bibitem [{\citenamefont {Schlawin}\ \emph {et~al.}(2019)\citenamefont
  {Schlawin}, \citenamefont {Cavalleri},\ and\ \citenamefont
  {Jaksch}}]{Frank1}%
  \BibitemOpen
  \bibfield  {author} {\bibinfo {author} {\bibfnamefont {F.}~\bibnamefont
  {Schlawin}}, \bibinfo {author} {\bibfnamefont {A.}~\bibnamefont {Cavalleri}},
  \ and\ \bibinfo {author} {\bibfnamefont {D.}~\bibnamefont {Jaksch}},\ }\href
  {\doibase 10.1103/PhysRevLett.122.133602} {\bibfield  {journal} {\bibinfo
  {journal} {Phys. Rev. Lett.}\ }\textbf {\bibinfo {volume} {122}},\ \bibinfo
  {pages} {133602} (\bibinfo {year} {2019})}\BibitemShut {NoStop}%
\bibitem [{\citenamefont {Fröml}\ \emph {et~al.}(2019)\citenamefont {Fröml},
  \citenamefont {Muckel}, \citenamefont {Kollath}, \citenamefont
  {Chiocchetta},\ and\ \citenamefont {Diehl}}]{KollathQuantumWire}%
  \BibitemOpen
  \bibfield  {author} {\bibinfo {author} {\bibfnamefont {H.}~\bibnamefont
  {Fröml}}, \bibinfo {author} {\bibfnamefont {C.}~\bibnamefont {Muckel}},
  \bibinfo {author} {\bibfnamefont {C.}~\bibnamefont {Kollath}}, \bibinfo
  {author} {\bibfnamefont {A.}~\bibnamefont {Chiocchetta}}, \ and\ \bibinfo
  {author} {\bibfnamefont {S.}~\bibnamefont {Diehl}},\ }\href@noop {} {\enquote
  {\bibinfo {title} {Ultracold quantum wires with localized losses: many-body
  quantum zeno effect},}\ } (\bibinfo {year} {2019}),\ \Eprint
  {http://arxiv.org/abs/1910.10741} {arXiv:1910.10741 [cond-mat.quant-gas]}
  \BibitemShut {NoStop}%
\bibitem [{\citenamefont {Lebrat}\ \emph {et~al.}(2019)\citenamefont {Lebrat},
  \citenamefont {H\"ausler}, \citenamefont {Fabritius}, \citenamefont
  {Husmann}, \citenamefont {Corman},\ and\ \citenamefont
  {Esslinger}}]{Esslingerlocal1}%
  \BibitemOpen
  \bibfield  {author} {\bibinfo {author} {\bibfnamefont {M.}~\bibnamefont
  {Lebrat}}, \bibinfo {author} {\bibfnamefont {S.}~\bibnamefont {H\"ausler}},
  \bibinfo {author} {\bibfnamefont {P.}~\bibnamefont {Fabritius}}, \bibinfo
  {author} {\bibfnamefont {D.}~\bibnamefont {Husmann}}, \bibinfo {author}
  {\bibfnamefont {L.}~\bibnamefont {Corman}}, \ and\ \bibinfo {author}
  {\bibfnamefont {T.}~\bibnamefont {Esslinger}},\ }\href {\doibase
  10.1103/PhysRevLett.123.193605} {\bibfield  {journal} {\bibinfo  {journal}
  {Phys. Rev. Lett.}\ }\textbf {\bibinfo {volume} {123}},\ \bibinfo {pages}
  {193605} (\bibinfo {year} {2019})}\BibitemShut {NoStop}%
\bibitem [{\citenamefont {Mitrano}\ \emph {et~al.}(2014)\citenamefont
  {Mitrano}, \citenamefont {Cotugno}, \citenamefont {Clark}, \citenamefont
  {Singla}, \citenamefont {Kaiser}, \citenamefont {St\"ahler}, \citenamefont
  {Beyer}, \citenamefont {Dressel}, \citenamefont {Baldassarre}, \citenamefont
  {Nicoletti}, \citenamefont {Perucchi}, \citenamefont {Hasegawa},
  \citenamefont {Okamoto}, \citenamefont {Jaksch},\ and\ \citenamefont
  {Cavalleri}}]{Andrea1}%
  \BibitemOpen
  \bibfield  {author} {\bibinfo {author} {\bibfnamefont {M.}~\bibnamefont
  {Mitrano}}, \bibinfo {author} {\bibfnamefont {G.}~\bibnamefont {Cotugno}},
  \bibinfo {author} {\bibfnamefont {S.~R.}\ \bibnamefont {Clark}}, \bibinfo
  {author} {\bibfnamefont {R.}~\bibnamefont {Singla}}, \bibinfo {author}
  {\bibfnamefont {S.}~\bibnamefont {Kaiser}}, \bibinfo {author} {\bibfnamefont
  {J.}~\bibnamefont {St\"ahler}}, \bibinfo {author} {\bibfnamefont
  {R.}~\bibnamefont {Beyer}}, \bibinfo {author} {\bibfnamefont
  {M.}~\bibnamefont {Dressel}}, \bibinfo {author} {\bibfnamefont
  {L.}~\bibnamefont {Baldassarre}}, \bibinfo {author} {\bibfnamefont
  {D.}~\bibnamefont {Nicoletti}}, \bibinfo {author} {\bibfnamefont
  {A.}~\bibnamefont {Perucchi}}, \bibinfo {author} {\bibfnamefont
  {T.}~\bibnamefont {Hasegawa}}, \bibinfo {author} {\bibfnamefont
  {H.}~\bibnamefont {Okamoto}}, \bibinfo {author} {\bibfnamefont
  {D.}~\bibnamefont {Jaksch}}, \ and\ \bibinfo {author} {\bibfnamefont
  {A.}~\bibnamefont {Cavalleri}},\ }\href {\doibase
  10.1103/PhysRevLett.112.117801} {\bibfield  {journal} {\bibinfo  {journal}
  {Phys. Rev. Lett.}\ }\textbf {\bibinfo {volume} {112}},\ \bibinfo {pages}
  {117801} (\bibinfo {year} {2014})}\BibitemShut {NoStop}%
\bibitem [{\citenamefont {De~Franceschi}\ \emph {et~al.}(2010)\citenamefont
  {De~Franceschi}, \citenamefont {Kouwenhoven}, \citenamefont
  {Sch{\"o}nenberger},\ and\ \citenamefont {Wernsdorfer}}]{solidstate}%
  \BibitemOpen
  \bibfield  {author} {\bibinfo {author} {\bibfnamefont {S.}~\bibnamefont
  {De~Franceschi}}, \bibinfo {author} {\bibfnamefont {L.}~\bibnamefont
  {Kouwenhoven}}, \bibinfo {author} {\bibfnamefont {C.}~\bibnamefont
  {Sch{\"o}nenberger}}, \ and\ \bibinfo {author} {\bibfnamefont
  {W.}~\bibnamefont {Wernsdorfer}},\ }\href {\doibase 10.1038/nnano.2010.173}
  {\bibfield  {journal} {\bibinfo  {journal} {Nature Nanotechnology}\ }\textbf
  {\bibinfo {volume} {5}},\ \bibinfo {pages} {703} (\bibinfo {year}
  {2010})}\BibitemShut {NoStop}%
\bibitem [{\citenamefont {Prosen}(2011{\natexlab{a}})}]{Prosen1}%
  \BibitemOpen
  \bibfield  {author} {\bibinfo {author} {\bibfnamefont {T.}~\bibnamefont
  {Prosen}},\ }\href {\doibase 10.1103/PhysRevLett.106.217206} {\bibfield
  {journal} {\bibinfo  {journal} {Phys. Rev. Lett.}\ }\textbf {\bibinfo
  {volume} {106}},\ \bibinfo {pages} {217206} (\bibinfo {year}
  {2011}{\natexlab{a}})}\BibitemShut {NoStop}%
\bibitem [{\citenamefont {Prosen}(2011{\natexlab{b}})}]{Prosen2}%
  \BibitemOpen
  \bibfield  {author} {\bibinfo {author} {\bibfnamefont {T.}~\bibnamefont
  {Prosen}},\ }\href {\doibase 10.1103/PhysRevLett.107.137201} {\bibfield
  {journal} {\bibinfo  {journal} {Phys. Rev. Lett.}\ }\textbf {\bibinfo
  {volume} {107}},\ \bibinfo {pages} {137201} (\bibinfo {year}
  {2011}{\natexlab{b}})}\BibitemShut {NoStop}%
\bibitem [{\citenamefont {Popkov}\ and\ \citenamefont
  {Prosen}(2015)}]{Prosen3}%
  \BibitemOpen
  \bibfield  {author} {\bibinfo {author} {\bibfnamefont {V.}~\bibnamefont
  {Popkov}}\ and\ \bibinfo {author} {\bibfnamefont {T.}~\bibnamefont
  {Prosen}},\ }\href {\doibase 10.1103/PhysRevLett.114.127201} {\bibfield
  {journal} {\bibinfo  {journal} {Phys. Rev. Lett.}\ }\textbf {\bibinfo
  {volume} {114}},\ \bibinfo {pages} {127201} (\bibinfo {year}
  {2015})}\BibitemShut {NoStop}%
\bibitem [{\citenamefont {Ilievski}(2014)}]{Enej1}%
  \BibitemOpen
  \bibfield  {author} {\bibinfo {author} {\bibfnamefont {E.}~\bibnamefont
  {Ilievski}},\ }\href@noop {} {\enquote {\bibinfo {title} {Exact solutions of
  open integrable quantum spin chains},}\ } (\bibinfo {year} {2014}),\ \Eprint
  {http://arxiv.org/abs/1410.1446} {arXiv:1410.1446 [quant-ph]} \BibitemShut
  {NoStop}%
\bibitem [{\citenamefont {Ilievski}(2017)}]{Enej2}%
  \BibitemOpen
  \bibfield  {author} {\bibinfo {author} {\bibfnamefont {E.}~\bibnamefont
  {Ilievski}},\ }\href {\doibase 10.21468/SciPostPhys.3.4.031} {\bibfield
  {journal} {\bibinfo  {journal} {SciPost Phys.}\ }\textbf {\bibinfo {volume}
  {3}},\ \bibinfo {pages} {031} (\bibinfo {year} {2017})}\BibitemShut {NoStop}%
\bibitem [{\citenamefont {Ilievski}\ and\ \citenamefont
  {Prosen}(2014)}]{Enej3}%
  \BibitemOpen
  \bibfield  {author} {\bibinfo {author} {\bibfnamefont {E.}~\bibnamefont
  {Ilievski}}\ and\ \bibinfo {author} {\bibfnamefont {T.}~\bibnamefont
  {Prosen}},\ }\href {\doibase https://doi.org/10.1016/j.nuclphysb.2014.03.016}
  {\bibfield  {journal} {\bibinfo  {journal} {Nuclear Physics B}\ }\textbf
  {\bibinfo {volume} {882}},\ \bibinfo {pages} {485 } (\bibinfo {year}
  {2014})}\BibitemShut {NoStop}%
\bibitem [{\citenamefont {Ilievski}\ and\ \citenamefont
  {{\v{Z}}unkovi{\v{c}}}(2014)}]{Bojan1}%
  \BibitemOpen
  \bibfield  {author} {\bibinfo {author} {\bibfnamefont {E.}~\bibnamefont
  {Ilievski}}\ and\ \bibinfo {author} {\bibfnamefont {B.}~\bibnamefont
  {{\v{Z}}unkovi{\v{c}}}},\ }\href {\doibase 10.1088/1742-5468/2014/01/p01001}
  {\bibfield  {journal} {\bibinfo  {journal} {Journal of Statistical Mechanics:
  Theory and Experiment}\ }\textbf {\bibinfo {volume} {2014}},\ \bibinfo
  {pages} {P01001} (\bibinfo {year} {2014})}\BibitemShut {NoStop}%
\bibitem [{\citenamefont {{\v{Z}}unkovi{\v{c}}}(2014)}]{Bojan2}%
  \BibitemOpen
  \bibfield  {author} {\bibinfo {author} {\bibfnamefont {B.}~\bibnamefont
  {{\v{Z}}unkovi{\v{c}}}},\ }\href {\doibase 10.1088/1367-2630/16/1/013042}
  {\bibfield  {journal} {\bibinfo  {journal} {New Journal of Physics}\ }\textbf
  {\bibinfo {volume} {16}},\ \bibinfo {pages} {013042} (\bibinfo {year}
  {2014})}\BibitemShut {NoStop}%
\bibitem [{\citenamefont {Lenar\ifmmode \check{c}\else
  \v{c}\fi{}i\ifmmode~\check{c}\else \v{c}\fi{}}\ and\ \citenamefont
  {Prosen}(2015)}]{Zala1}%
  \BibitemOpen
  \bibfield  {author} {\bibinfo {author} {\bibfnamefont {Z.}~\bibnamefont
  {Lenar\ifmmode \check{c}\else \v{c}\fi{}i\ifmmode~\check{c}\else
  \v{c}\fi{}}}\ and\ \bibinfo {author} {\bibfnamefont {T.}~\bibnamefont
  {Prosen}},\ }\href {\doibase 10.1103/PhysRevE.91.030103} {\bibfield
  {journal} {\bibinfo  {journal} {Phys. Rev. E}\ }\textbf {\bibinfo {volume}
  {91}},\ \bibinfo {pages} {030103} (\bibinfo {year} {2015})}\BibitemShut
  {NoStop}%
\bibitem [{\citenamefont {Karevski}\ \emph {et~al.}(2013)\citenamefont
  {Karevski}, \citenamefont {Popkov},\ and\ \citenamefont
  {Sch\"utz}}]{Popkov1}%
  \BibitemOpen
  \bibfield  {author} {\bibinfo {author} {\bibfnamefont {D.}~\bibnamefont
  {Karevski}}, \bibinfo {author} {\bibfnamefont {V.}~\bibnamefont {Popkov}}, \
  and\ \bibinfo {author} {\bibfnamefont {G.~M.}\ \bibnamefont {Sch\"utz}},\
  }\href {\doibase 10.1103/PhysRevLett.110.047201} {\bibfield  {journal}
  {\bibinfo  {journal} {Phys. Rev. Lett.}\ }\textbf {\bibinfo {volume} {110}},\
  \bibinfo {pages} {047201} (\bibinfo {year} {2013})}\BibitemShut {NoStop}%
\bibitem [{\citenamefont {Popkov}\ and\ \citenamefont
  {Sch\"utz}(2017)}]{Popkov2}%
  \BibitemOpen
  \bibfield  {author} {\bibinfo {author} {\bibfnamefont {V.}~\bibnamefont
  {Popkov}}\ and\ \bibinfo {author} {\bibfnamefont {G.~M.}\ \bibnamefont
  {Sch\"utz}},\ }\href {\doibase 10.1103/PhysRevE.95.042128} {\bibfield
  {journal} {\bibinfo  {journal} {Phys. Rev. E}\ }\textbf {\bibinfo {volume}
  {95}},\ \bibinfo {pages} {042128} (\bibinfo {year} {2017})}\BibitemShut
  {NoStop}%
\bibitem [{\citenamefont {Yuge}\ and\ \citenamefont {Sugita}(2015)}]{Yuge}%
  \BibitemOpen
  \bibfield  {author} {\bibinfo {author} {\bibfnamefont {T.}~\bibnamefont
  {Yuge}}\ and\ \bibinfo {author} {\bibfnamefont {A.}~\bibnamefont {Sugita}},\
  }\href@noop {} {\bibfield  {journal} {\bibinfo  {journal} {Journal of the
  Physical Society of Japan}\ }\textbf {\bibinfo {volume} {84}},\ \bibinfo
  {pages} {014001} (\bibinfo {year} {2015})}\BibitemShut {NoStop}%
\bibitem [{\citenamefont {Popkov}\ \emph {et~al.}(2019)\citenamefont {Popkov},
  \citenamefont {Prosen},\ and\ \citenamefont {Zadnik}}]{Lenart1}%
  \BibitemOpen
  \bibfield  {author} {\bibinfo {author} {\bibfnamefont {V.}~\bibnamefont
  {Popkov}}, \bibinfo {author} {\bibfnamefont {T.}~\bibnamefont {Prosen}}, \
  and\ \bibinfo {author} {\bibfnamefont {L.}~\bibnamefont {Zadnik}},\
  }\href@noop {} {\enquote {\bibinfo {title} {Exact nonequilibrium steady state
  of open xxz/xyz spin-1/2 chain with dirichlet boundary conditions},}\ }
  (\bibinfo {year} {2019}),\ \Eprint {http://arxiv.org/abs/1912.03282}
  {arXiv:1912.03282 [cond-mat.stat-mech]} \BibitemShut {NoStop}%
\bibitem [{\citenamefont {Vanicat}\ \emph {et~al.}(2018)\citenamefont
  {Vanicat}, \citenamefont {Zadnik},\ and\ \citenamefont {Prosen}}]{Lenart2}%
  \BibitemOpen
  \bibfield  {author} {\bibinfo {author} {\bibfnamefont {M.}~\bibnamefont
  {Vanicat}}, \bibinfo {author} {\bibfnamefont {L.}~\bibnamefont {Zadnik}}, \
  and\ \bibinfo {author} {\bibfnamefont {T.}~\bibnamefont {Prosen}},\ }\href
  {\doibase 10.1103/PhysRevLett.121.030606} {\bibfield  {journal} {\bibinfo
  {journal} {Phys. Rev. Lett.}\ }\textbf {\bibinfo {volume} {121}},\ \bibinfo
  {pages} {030606} (\bibinfo {year} {2018})}\BibitemShut {NoStop}%
\bibitem [{\citenamefont {Bu\ifmmode~\check{c}\else \v{c}\fi{}a}\ and\
  \citenamefont {Prosen}(2014)}]{BucaProsen1}%
  \BibitemOpen
  \bibfield  {author} {\bibinfo {author} {\bibfnamefont {B.}~\bibnamefont
  {Bu\ifmmode~\check{c}\else \v{c}\fi{}a}}\ and\ \bibinfo {author}
  {\bibfnamefont {T.}~\bibnamefont {Prosen}},\ }\href {\doibase
  10.1103/PhysRevLett.112.067201} {\bibfield  {journal} {\bibinfo  {journal}
  {Phys. Rev. Lett.}\ }\textbf {\bibinfo {volume} {112}},\ \bibinfo {pages}
  {067201} (\bibinfo {year} {2014})}\BibitemShut {NoStop}%
\bibitem [{\citenamefont {Nigro}(2020)}]{Nigro}%
  \BibitemOpen
  \bibfield  {author} {\bibinfo {author} {\bibfnamefont {D.}~\bibnamefont
  {Nigro}},\ }\href {\doibase 10.1103/PhysRevA.101.022109} {\bibfield
  {journal} {\bibinfo  {journal} {Phys. Rev. A}\ }\textbf {\bibinfo {volume}
  {101}},\ \bibinfo {pages} {022109} (\bibinfo {year} {2020})}\BibitemShut
  {NoStop}%
\bibitem [{\citenamefont {Prosen}(2015)}]{ProsenReview}%
  \BibitemOpen
  \bibfield  {author} {\bibinfo {author} {\bibfnamefont {T.}~\bibnamefont
  {Prosen}},\ }\href {\doibase 10.1088/1751-8113/48/37/373001} {\bibfield
  {journal} {\bibinfo  {journal} {Journal of Physics A: Mathematical and
  Theoretical}\ }\textbf {\bibinfo {volume} {48}},\ \bibinfo {pages} {373001}
  (\bibinfo {year} {2015})}\BibitemShut {NoStop}%
\bibitem [{\citenamefont {Diehl}\ \emph {et~al.}(2008)\citenamefont {Diehl},
  \citenamefont {Micheli}, \citenamefont {Kantian}, \citenamefont {Kraus},
  \citenamefont {B{\"u}chler},\ and\ \citenamefont {Zoller}}]{Diehl1}%
  \BibitemOpen
  \bibfield  {author} {\bibinfo {author} {\bibfnamefont {S.}~\bibnamefont
  {Diehl}}, \bibinfo {author} {\bibfnamefont {A.}~\bibnamefont {Micheli}},
  \bibinfo {author} {\bibfnamefont {A.}~\bibnamefont {Kantian}}, \bibinfo
  {author} {\bibfnamefont {B.}~\bibnamefont {Kraus}}, \bibinfo {author}
  {\bibfnamefont {H.~P.}\ \bibnamefont {B{\"u}chler}}, \ and\ \bibinfo {author}
  {\bibfnamefont {P.}~\bibnamefont {Zoller}},\ }\href {\doibase
  10.1038/nphys1073} {\bibfield  {journal} {\bibinfo  {journal} {Nature
  Physics}\ }\textbf {\bibinfo {volume} {4}},\ \bibinfo {pages} {878} (\bibinfo
  {year} {2008})}\BibitemShut {NoStop}%
\bibitem [{\citenamefont {Bu{\v c}a}\ and\ \citenamefont
  {Prosen}(2018)}]{BucaProsen3}%
  \BibitemOpen
  \bibfield  {author} {\bibinfo {author} {\bibfnamefont {B.}~\bibnamefont
  {Bu{\v c}a}}\ and\ \bibinfo {author} {\bibfnamefont {T.}~\bibnamefont
  {Prosen}},\ }\href {\doibase 10.1140/epjst/e2018-00100-9} {\bibfield
  {journal} {\bibinfo  {journal} {The European Physical Journal Special
  Topics}\ }\textbf {\bibinfo {volume} {227}},\ \bibinfo {pages} {421}
  (\bibinfo {year} {2018})}\BibitemShut {NoStop}%
\bibitem [{\citenamefont {\v{Z}nidari\ifmmode~\check{c}\else
  \v{c}\fi{}}(2011)}]{Znidaric1}%
  \BibitemOpen
  \bibfield  {author} {\bibinfo {author} {\bibfnamefont {M.}~\bibnamefont
  {\v{Z}nidari\ifmmode~\check{c}\else \v{c}\fi{}}},\ }\href {\doibase
  10.1103/PhysRevLett.106.220601} {\bibfield  {journal} {\bibinfo  {journal}
  {Phys. Rev. Lett.}\ }\textbf {\bibinfo {volume} {106}},\ \bibinfo {pages}
  {220601} (\bibinfo {year} {2011})}\BibitemShut {NoStop}%
\bibitem [{\citenamefont {\v{Z}nidari\ifmmode~\check{c}\else
  \v{c}\fi{}}(2014)}]{Znidaric2}%
  \BibitemOpen
  \bibfield  {author} {\bibinfo {author} {\bibfnamefont {M.}~\bibnamefont
  {\v{Z}nidari\ifmmode~\check{c}\else \v{c}\fi{}}},\ }\href {\doibase
  10.1103/PhysRevLett.112.040602} {\bibfield  {journal} {\bibinfo  {journal}
  {Phys. Rev. Lett.}\ }\textbf {\bibinfo {volume} {112}},\ \bibinfo {pages}
  {040602} (\bibinfo {year} {2014})}\BibitemShut {NoStop}%
\bibitem [{\citenamefont {{\v{Z}}nidari{\v{c}}}(2010)}]{Znidaric3}%
  \BibitemOpen
  \bibfield  {author} {\bibinfo {author} {\bibfnamefont {M.}~\bibnamefont
  {{\v{Z}}nidari{\v{c}}}},\ }\href {\doibase 10.1088/1742-5468/2010/05/l05002}
  {\bibfield  {journal} {\bibinfo  {journal} {Journal of Statistical Mechanics:
  Theory and Experiment}\ }\textbf {\bibinfo {volume} {2010}},\ \bibinfo
  {pages} {L05002} (\bibinfo {year} {2010})}\BibitemShut {NoStop}%
\bibitem [{\citenamefont {Prosen}(2008)}]{Prosen3rd1}%
  \BibitemOpen
  \bibfield  {author} {\bibinfo {author} {\bibfnamefont {T.}~\bibnamefont
  {Prosen}},\ }\href {\doibase 10.1088/1367-2630/10/4/043026} {\bibfield
  {journal} {\bibinfo  {journal} {New Journal of Physics}\ }\textbf {\bibinfo
  {volume} {10}},\ \bibinfo {pages} {043026} (\bibinfo {year}
  {2008})}\BibitemShut {NoStop}%
\bibitem [{\citenamefont {Prosen}\ and\ \citenamefont
  {Seligman}(2010)}]{Prosen3rd2}%
  \BibitemOpen
  \bibfield  {author} {\bibinfo {author} {\bibfnamefont {T.}~\bibnamefont
  {Prosen}}\ and\ \bibinfo {author} {\bibfnamefont {T.~H.}\ \bibnamefont
  {Seligman}},\ }\href {\doibase 10.1088/1751-8113/43/39/392004} {\bibfield
  {journal} {\bibinfo  {journal} {Journal of Physics A: Mathematical and
  Theoretical}\ }\textbf {\bibinfo {volume} {43}},\ \bibinfo {pages} {392004}
  (\bibinfo {year} {2010})}\BibitemShut {NoStop}%
\bibitem [{\citenamefont {Manzano}\ \emph {et~al.}(2016)\citenamefont
  {Manzano}, \citenamefont {Chuang},\ and\ \citenamefont {Cao}}]{Manzano1}%
  \BibitemOpen
  \bibfield  {author} {\bibinfo {author} {\bibfnamefont {D.}~\bibnamefont
  {Manzano}}, \bibinfo {author} {\bibfnamefont {C.}~\bibnamefont {Chuang}}, \
  and\ \bibinfo {author} {\bibfnamefont {J.}~\bibnamefont {Cao}},\ }\href
  {\doibase 10.1088/1367-2630/18/4/043044} {\bibfield  {journal} {\bibinfo
  {journal} {New Journal of Physics}\ }\textbf {\bibinfo {volume} {18}},\
  \bibinfo {pages} {043044} (\bibinfo {year} {2016})}\BibitemShut {NoStop}%
\bibitem [{\citenamefont {Monthus}(2017)}]{Monthus1}%
  \BibitemOpen
  \bibfield  {author} {\bibinfo {author} {\bibfnamefont {C.}~\bibnamefont
  {Monthus}},\ }\href {\doibase 10.1088/1742-5468/aa6a2f} {\bibfield  {journal}
  {\bibinfo  {journal} {Journal of Statistical Mechanics: Theory and
  Experiment}\ }\textbf {\bibinfo {volume} {2017}},\ \bibinfo {pages} {043303}
  (\bibinfo {year} {2017})}\BibitemShut {NoStop}%
\bibitem [{\citenamefont {Krapivsky}\ \emph {et~al.}(2019)\citenamefont
  {Krapivsky}, \citenamefont {Mallick},\ and\ \citenamefont
  {Sels}}]{Krapivsky}%
  \BibitemOpen
  \bibfield  {author} {\bibinfo {author} {\bibfnamefont {P.~L.}\ \bibnamefont
  {Krapivsky}}, \bibinfo {author} {\bibfnamefont {K.}~\bibnamefont {Mallick}},
  \ and\ \bibinfo {author} {\bibfnamefont {D.}~\bibnamefont {Sels}},\ }\href
  {\doibase 10.1088/1742-5468/ab4e8e} {\bibfield  {journal} {\bibinfo
  {journal} {Journal of Statistical Mechanics: Theory and Experiment}\ }\textbf
  {\bibinfo {volume} {2019}},\ \bibinfo {pages} {113108} (\bibinfo {year}
  {2019})}\BibitemShut {NoStop}%
\bibitem [{\citenamefont {Carollo}\ \emph {et~al.}(2017)\citenamefont
  {Carollo}, \citenamefont {Garrahan}, \citenamefont {Lesanovsky},\ and\
  \citenamefont {P\'erez-Espigares}}]{JuanP1}%
  \BibitemOpen
  \bibfield  {author} {\bibinfo {author} {\bibfnamefont {F.}~\bibnamefont
  {Carollo}}, \bibinfo {author} {\bibfnamefont {J.~P.}\ \bibnamefont
  {Garrahan}}, \bibinfo {author} {\bibfnamefont {I.}~\bibnamefont
  {Lesanovsky}}, \ and\ \bibinfo {author} {\bibfnamefont {C.}~\bibnamefont
  {P\'erez-Espigares}},\ }\href {\doibase 10.1103/PhysRevE.96.052118}
  {\bibfield  {journal} {\bibinfo  {journal} {Phys. Rev. E}\ }\textbf {\bibinfo
  {volume} {96}},\ \bibinfo {pages} {052118} (\bibinfo {year}
  {2017})}\BibitemShut {NoStop}%
\bibitem [{\citenamefont {Budich}\ \emph {et~al.}(2015)\citenamefont {Budich},
  \citenamefont {Zoller},\ and\ \citenamefont {Diehl}}]{Zoller}%
  \BibitemOpen
  \bibfield  {author} {\bibinfo {author} {\bibfnamefont {J.~C.}\ \bibnamefont
  {Budich}}, \bibinfo {author} {\bibfnamefont {P.}~\bibnamefont {Zoller}}, \
  and\ \bibinfo {author} {\bibfnamefont {S.}~\bibnamefont {Diehl}},\ }\href
  {\doibase 10.1103/PhysRevA.91.042117} {\bibfield  {journal} {\bibinfo
  {journal} {Phys. Rev. A}\ }\textbf {\bibinfo {volume} {91}},\ \bibinfo
  {pages} {042117} (\bibinfo {year} {2015})}\BibitemShut {NoStop}%
\bibitem [{\citenamefont {Iemini}\ \emph {et~al.}(2016)\citenamefont {Iemini},
  \citenamefont {Rossini}, \citenamefont {Fazio}, \citenamefont {Diehl},\ and\
  \citenamefont {Mazza}}]{Diehl2}%
  \BibitemOpen
  \bibfield  {author} {\bibinfo {author} {\bibfnamefont {F.}~\bibnamefont
  {Iemini}}, \bibinfo {author} {\bibfnamefont {D.}~\bibnamefont {Rossini}},
  \bibinfo {author} {\bibfnamefont {R.}~\bibnamefont {Fazio}}, \bibinfo
  {author} {\bibfnamefont {S.}~\bibnamefont {Diehl}}, \ and\ \bibinfo {author}
  {\bibfnamefont {L.}~\bibnamefont {Mazza}},\ }\href {\doibase
  10.1103/PhysRevB.93.115113} {\bibfield  {journal} {\bibinfo  {journal} {Phys.
  Rev. B}\ }\textbf {\bibinfo {volume} {93}},\ \bibinfo {pages} {115113}
  (\bibinfo {year} {2016})}\BibitemShut {NoStop}%
\bibitem [{\citenamefont {Medvedyeva}\ and\ \citenamefont
  {Kehrein}(2014)}]{Mariya1}%
  \BibitemOpen
  \bibfield  {author} {\bibinfo {author} {\bibfnamefont {M.~V.}\ \bibnamefont
  {Medvedyeva}}\ and\ \bibinfo {author} {\bibfnamefont {S.}~\bibnamefont
  {Kehrein}},\ }\href {\doibase 10.1103/PhysRevB.90.205410} {\bibfield
  {journal} {\bibinfo  {journal} {Phys. Rev. B}\ }\textbf {\bibinfo {volume}
  {90}},\ \bibinfo {pages} {205410} (\bibinfo {year} {2014})}\BibitemShut
  {NoStop}%
\bibitem [{\citenamefont {Guo}\ and\ \citenamefont {Poletti}(2017)}]{Guo}%
  \BibitemOpen
  \bibfield  {author} {\bibinfo {author} {\bibfnamefont {C.}~\bibnamefont
  {Guo}}\ and\ \bibinfo {author} {\bibfnamefont {D.}~\bibnamefont {Poletti}},\
  }\href {\doibase 10.1103/PhysRevA.95.052107} {\bibfield  {journal} {\bibinfo
  {journal} {Phys. Rev. A}\ }\textbf {\bibinfo {volume} {95}},\ \bibinfo
  {pages} {052107} (\bibinfo {year} {2017})}\BibitemShut {NoStop}%
\bibitem [{\citenamefont {Medvedyeva}\ \emph {et~al.}(2016)\citenamefont
  {Medvedyeva}, \citenamefont {Essler},\ and\ \citenamefont
  {Prosen}}]{Essler1}%
  \BibitemOpen
  \bibfield  {author} {\bibinfo {author} {\bibfnamefont {M.~V.}\ \bibnamefont
  {Medvedyeva}}, \bibinfo {author} {\bibfnamefont {F.~H.~L.}\ \bibnamefont
  {Essler}}, \ and\ \bibinfo {author} {\bibfnamefont {T.~c.~v.}\ \bibnamefont
  {Prosen}},\ }\href {\doibase 10.1103/PhysRevLett.117.137202} {\bibfield
  {journal} {\bibinfo  {journal} {Phys. Rev. Lett.}\ }\textbf {\bibinfo
  {volume} {117}},\ \bibinfo {pages} {137202} (\bibinfo {year}
  {2016})}\BibitemShut {NoStop}%
\bibitem [{\citenamefont {Rowlands}\ and\ \citenamefont
  {Lamacraft}(2018)}]{Lamacraft}%
  \BibitemOpen
  \bibfield  {author} {\bibinfo {author} {\bibfnamefont {D.~A.}\ \bibnamefont
  {Rowlands}}\ and\ \bibinfo {author} {\bibfnamefont {A.}~\bibnamefont
  {Lamacraft}},\ }\href {\doibase 10.1103/PhysRevLett.120.090401} {\bibfield
  {journal} {\bibinfo  {journal} {Phys. Rev. Lett.}\ }\textbf {\bibinfo
  {volume} {120}},\ \bibinfo {pages} {090401} (\bibinfo {year}
  {2018})}\BibitemShut {NoStop}%
\bibitem [{\citenamefont {Ziolkowska}\ and\ \citenamefont
  {Essler}(2020)}]{Essler2}%
  \BibitemOpen
  \bibfield  {author} {\bibinfo {author} {\bibfnamefont {A.~A.}\ \bibnamefont
  {Ziolkowska}}\ and\ \bibinfo {author} {\bibfnamefont {F.~H.}\ \bibnamefont
  {Essler}},\ }\href {\doibase 10.21468/SciPostPhys.8.3.044} {\bibfield
  {journal} {\bibinfo  {journal} {SciPost Phys.}\ }\textbf {\bibinfo {volume}
  {8}},\ \bibinfo {pages} {44} (\bibinfo {year} {2020})}\BibitemShut {NoStop}%
\bibitem [{\citenamefont {Shibata}\ and\ \citenamefont
  {Katsura}(2019{\natexlab{a}})}]{Katsura1}%
  \BibitemOpen
  \bibfield  {author} {\bibinfo {author} {\bibfnamefont {N.}~\bibnamefont
  {Shibata}}\ and\ \bibinfo {author} {\bibfnamefont {H.}~\bibnamefont
  {Katsura}},\ }\href {\doibase 10.1103/PhysRevB.99.224432} {\bibfield
  {journal} {\bibinfo  {journal} {Phys. Rev. B}\ }\textbf {\bibinfo {volume}
  {99}},\ \bibinfo {pages} {224432} (\bibinfo {year}
  {2019}{\natexlab{a}})}\BibitemShut {NoStop}%
\bibitem [{\citenamefont {Shibata}\ and\ \citenamefont
  {Katsura}(2019{\natexlab{b}})}]{Katsura2}%
  \BibitemOpen
  \bibfield  {author} {\bibinfo {author} {\bibfnamefont {N.}~\bibnamefont
  {Shibata}}\ and\ \bibinfo {author} {\bibfnamefont {H.}~\bibnamefont
  {Katsura}},\ }\href {\doibase 10.1103/PhysRevB.99.174303} {\bibfield
  {journal} {\bibinfo  {journal} {Phys. Rev. B}\ }\textbf {\bibinfo {volume}
  {99}},\ \bibinfo {pages} {174303} (\bibinfo {year}
  {2019}{\natexlab{b}})}\BibitemShut {NoStop}%
\bibitem [{\citenamefont {van Caspel}\ and\ \citenamefont
  {Gritsev}(2018)}]{Gritsev}%
  \BibitemOpen
  \bibfield  {author} {\bibinfo {author} {\bibfnamefont {M.}~\bibnamefont {van
  Caspel}}\ and\ \bibinfo {author} {\bibfnamefont {V.}~\bibnamefont
  {Gritsev}},\ }\href {\doibase 10.1103/PhysRevA.97.052106} {\bibfield
  {journal} {\bibinfo  {journal} {Phys. Rev. A}\ }\textbf {\bibinfo {volume}
  {97}},\ \bibinfo {pages} {052106} (\bibinfo {year} {2018})}\BibitemShut
  {NoStop}%
\bibitem [{\citenamefont {Bastianello}\ \emph {et~al.}(2020)\citenamefont
  {Bastianello}, \citenamefont {Nardis},\ and\ \citenamefont
  {Luca}}]{preprint_hydro}%
  \BibitemOpen
  \bibfield  {author} {\bibinfo {author} {\bibfnamefont {A.}~\bibnamefont
  {Bastianello}}, \bibinfo {author} {\bibfnamefont {J.~D.}\ \bibnamefont
  {Nardis}}, \ and\ \bibinfo {author} {\bibfnamefont {A.~D.}\ \bibnamefont
  {Luca}},\ }\href@noop {} {\enquote {\bibinfo {title} {Generalised
  hydrodynamics with dephasing noise},}\ } (\bibinfo {year} {2020}),\ \Eprint
  {http://arxiv.org/abs/2003.01702} {arXiv:2003.01702 [cond-mat.stat-mech]}
  \BibitemShut {NoStop}%
\bibitem [{\citenamefont {Lange}\ \emph {et~al.}(2017)\citenamefont {Lange},
  \citenamefont {Lenar{\v c}i{\v c}},\ and\ \citenamefont {Rosch}}]{Zala2}%
  \BibitemOpen
  \bibfield  {author} {\bibinfo {author} {\bibfnamefont {F.}~\bibnamefont
  {Lange}}, \bibinfo {author} {\bibfnamefont {Z.}~\bibnamefont {Lenar{\v c}i{\v
  c}}}, \ and\ \bibinfo {author} {\bibfnamefont {A.}~\bibnamefont {Rosch}},\
  }\href {\doibase 10.1038/ncomms15767} {\bibfield  {journal} {\bibinfo
  {journal} {Nature Communications}\ }\textbf {\bibinfo {volume} {8}},\
  \bibinfo {pages} {15767} (\bibinfo {year} {2017})}\BibitemShut {NoStop}%
\bibitem [{\citenamefont {Lenar\ifmmode \check{c}\else
  \v{c}\fi{}i\ifmmode~\check{c}\else \v{c}\fi{}}\ \emph
  {et~al.}(2018)\citenamefont {Lenar\ifmmode \check{c}\else
  \v{c}\fi{}i\ifmmode~\check{c}\else \v{c}\fi{}}, \citenamefont {Lange},\ and\
  \citenamefont {Rosch}}]{Zala3}%
  \BibitemOpen
  \bibfield  {author} {\bibinfo {author} {\bibfnamefont {Z.}~\bibnamefont
  {Lenar\ifmmode \check{c}\else \v{c}\fi{}i\ifmmode~\check{c}\else
  \v{c}\fi{}}}, \bibinfo {author} {\bibfnamefont {F.}~\bibnamefont {Lange}}, \
  and\ \bibinfo {author} {\bibfnamefont {A.}~\bibnamefont {Rosch}},\ }\href
  {\doibase 10.1103/PhysRevB.97.024302} {\bibfield  {journal} {\bibinfo
  {journal} {Phys. Rev. B}\ }\textbf {\bibinfo {volume} {97}},\ \bibinfo
  {pages} {024302} (\bibinfo {year} {2018})}\BibitemShut {NoStop}%
\bibitem [{\citenamefont {Tonielli}\ \emph {et~al.}(2019)\citenamefont
  {Tonielli}, \citenamefont {Fazio}, \citenamefont {Diehl},\ and\ \citenamefont
  {Marino}}]{Ortocat}%
  \BibitemOpen
  \bibfield  {author} {\bibinfo {author} {\bibfnamefont {F.}~\bibnamefont
  {Tonielli}}, \bibinfo {author} {\bibfnamefont {R.}~\bibnamefont {Fazio}},
  \bibinfo {author} {\bibfnamefont {S.}~\bibnamefont {Diehl}}, \ and\ \bibinfo
  {author} {\bibfnamefont {J.}~\bibnamefont {Marino}},\ }\href {\doibase
  10.1103/PhysRevLett.122.040604} {\bibfield  {journal} {\bibinfo  {journal}
  {Phys. Rev. Lett.}\ }\textbf {\bibinfo {volume} {122}},\ \bibinfo {pages}
  {040604} (\bibinfo {year} {2019})}\BibitemShut {NoStop}%
\bibitem [{\citenamefont {Kuhr}(2016)}]{Kuhr}%
  \BibitemOpen
  \bibfield  {author} {\bibinfo {author} {\bibfnamefont {S.}~\bibnamefont
  {Kuhr}},\ }\href {\doibase 10.1093/nsr/nww023} {\bibfield  {journal}
  {\bibinfo  {journal} {National Science Review}\ }\textbf {\bibinfo {volume}
  {3}},\ \bibinfo {pages} {170} (\bibinfo {year} {2016})}\BibitemShut {NoStop}%
\bibitem [{\citenamefont {Damanet}\ \emph {et~al.}(2019)\citenamefont
  {Damanet}, \citenamefont {Mascarenhas}, \citenamefont {Pekker},\ and\
  \citenamefont {Daley}}]{Daley1}%
  \BibitemOpen
  \bibfield  {author} {\bibinfo {author} {\bibfnamefont {F.}~\bibnamefont
  {Damanet}}, \bibinfo {author} {\bibfnamefont {E.}~\bibnamefont
  {Mascarenhas}}, \bibinfo {author} {\bibfnamefont {D.}~\bibnamefont {Pekker}},
  \ and\ \bibinfo {author} {\bibfnamefont {A.~J.}\ \bibnamefont {Daley}},\
  }\href {\doibase 10.1103/PhysRevLett.123.180402} {\bibfield  {journal}
  {\bibinfo  {journal} {Phys. Rev. Lett.}\ }\textbf {\bibinfo {volume} {123}},\
  \bibinfo {pages} {180402} (\bibinfo {year} {2019})}\BibitemShut {NoStop}%
\bibitem [{\citenamefont {Barontini}\ \emph {et~al.}(2013)\citenamefont
  {Barontini}, \citenamefont {Labouvie}, \citenamefont {Stubenrauch},
  \citenamefont {Vogler}, \citenamefont {Guarrera},\ and\ \citenamefont
  {Ott}}]{LocalBEC}%
  \BibitemOpen
  \bibfield  {author} {\bibinfo {author} {\bibfnamefont {G.}~\bibnamefont
  {Barontini}}, \bibinfo {author} {\bibfnamefont {R.}~\bibnamefont {Labouvie}},
  \bibinfo {author} {\bibfnamefont {F.}~\bibnamefont {Stubenrauch}}, \bibinfo
  {author} {\bibfnamefont {A.}~\bibnamefont {Vogler}}, \bibinfo {author}
  {\bibfnamefont {V.}~\bibnamefont {Guarrera}}, \ and\ \bibinfo {author}
  {\bibfnamefont {H.}~\bibnamefont {Ott}},\ }\href {\doibase
  10.1103/PhysRevLett.110.035302} {\bibfield  {journal} {\bibinfo  {journal}
  {Phys. Rev. Lett.}\ }\textbf {\bibinfo {volume} {110}},\ \bibinfo {pages}
  {035302} (\bibinfo {year} {2013})}\BibitemShut {NoStop}%
\bibitem [{\citenamefont {Wolff}\ \emph {et~al.}(2020)\citenamefont {Wolff},
  \citenamefont {Sheikhan}, \citenamefont {Diehl},\ and\ \citenamefont
  {Kollath}}]{KollathQuantumWire2}%
  \BibitemOpen
  \bibfield  {author} {\bibinfo {author} {\bibfnamefont {S.}~\bibnamefont
  {Wolff}}, \bibinfo {author} {\bibfnamefont {A.}~\bibnamefont {Sheikhan}},
  \bibinfo {author} {\bibfnamefont {S.}~\bibnamefont {Diehl}}, \ and\ \bibinfo
  {author} {\bibfnamefont {C.}~\bibnamefont {Kollath}},\ }\href {\doibase
  10.1103/PhysRevB.101.075139} {\bibfield  {journal} {\bibinfo  {journal}
  {Phys. Rev. B}\ }\textbf {\bibinfo {volume} {101}},\ \bibinfo {pages}
  {075139} (\bibinfo {year} {2020})}\BibitemShut {NoStop}%
\bibitem [{\citenamefont {Bu{\v{c}}a}\ and\ \citenamefont
  {Prosen}(2012)}]{BucaProsen}%
  \BibitemOpen
  \bibfield  {author} {\bibinfo {author} {\bibfnamefont {B.}~\bibnamefont
  {Bu{\v{c}}a}}\ and\ \bibinfo {author} {\bibfnamefont {T.}~\bibnamefont
  {Prosen}},\ }\href {\doibase 10.1088/1367-2630/14/7/073007} {\bibfield
  {journal} {\bibinfo  {journal} {New Journal of Physics}\ }\textbf {\bibinfo
  {volume} {14}},\ \bibinfo {pages} {073007} (\bibinfo {year}
  {2012})}\BibitemShut {NoStop}%
\bibitem [{\citenamefont {Bu\ifmmode~\check{c}\else \v{c}\fi{}a}\ and\
  \citenamefont {Prosen}(2017)}]{BucaProsen2}%
  \BibitemOpen
  \bibfield  {author} {\bibinfo {author} {\bibfnamefont {B.}~\bibnamefont
  {Bu\ifmmode~\check{c}\else \v{c}\fi{}a}}\ and\ \bibinfo {author}
  {\bibfnamefont {T.}~\bibnamefont {Prosen}},\ }\href {\doibase
  10.1103/PhysRevE.95.052141} {\bibfield  {journal} {\bibinfo  {journal} {Phys.
  Rev. E}\ }\textbf {\bibinfo {volume} {95}},\ \bibinfo {pages} {052141}
  (\bibinfo {year} {2017})}\BibitemShut {NoStop}%
\bibitem [{\citenamefont {Mendoza-Arenas}\ \emph {et~al.}(2014)\citenamefont
  {Mendoza-Arenas}, \citenamefont {Mitchison}, \citenamefont {Clark},
  \citenamefont {Prior}, \citenamefont {Jaksch},\ and\ \citenamefont
  {Plenio}}]{Mendoza1}%
  \BibitemOpen
  \bibfield  {author} {\bibinfo {author} {\bibfnamefont {J.~J.}\ \bibnamefont
  {Mendoza-Arenas}}, \bibinfo {author} {\bibfnamefont {M.~T.}\ \bibnamefont
  {Mitchison}}, \bibinfo {author} {\bibfnamefont {S.~R.}\ \bibnamefont
  {Clark}}, \bibinfo {author} {\bibfnamefont {J.}~\bibnamefont {Prior}},
  \bibinfo {author} {\bibfnamefont {D.}~\bibnamefont {Jaksch}}, \ and\ \bibinfo
  {author} {\bibfnamefont {M.~B.}\ \bibnamefont {Plenio}},\ }\href {\doibase
  10.1088/1367-2630/16/5/053016} {\bibfield  {journal} {\bibinfo  {journal}
  {New Journal of Physics}\ }\textbf {\bibinfo {volume} {16}},\ \bibinfo
  {pages} {053016} (\bibinfo {year} {2014})}\BibitemShut {NoStop}%
\bibitem [{\citenamefont {Horstmann}\ \emph {et~al.}(2013)\citenamefont
  {Horstmann}, \citenamefont {Cirac},\ and\ \citenamefont
  {Giedke}}]{CiracDPT2}%
  \BibitemOpen
  \bibfield  {author} {\bibinfo {author} {\bibfnamefont {B.}~\bibnamefont
  {Horstmann}}, \bibinfo {author} {\bibfnamefont {J.~I.}\ \bibnamefont
  {Cirac}}, \ and\ \bibinfo {author} {\bibfnamefont {G.}~\bibnamefont
  {Giedke}},\ }\href {\doibase 10.1103/PhysRevA.87.012108} {\bibfield
  {journal} {\bibinfo  {journal} {Phys. Rev. A}\ }\textbf {\bibinfo {volume}
  {87}},\ \bibinfo {pages} {012108} (\bibinfo {year} {2013})}\BibitemShut
  {NoStop}%
\bibitem [{\citenamefont {Kessler}\ \emph {et~al.}(2012)\citenamefont
  {Kessler}, \citenamefont {Giedke}, \citenamefont {Imamoglu}, \citenamefont
  {Yelin}, \citenamefont {Lukin},\ and\ \citenamefont {Cirac}}]{CiracDPT}%
  \BibitemOpen
  \bibfield  {author} {\bibinfo {author} {\bibfnamefont {E.~M.}\ \bibnamefont
  {Kessler}}, \bibinfo {author} {\bibfnamefont {G.}~\bibnamefont {Giedke}},
  \bibinfo {author} {\bibfnamefont {A.}~\bibnamefont {Imamoglu}}, \bibinfo
  {author} {\bibfnamefont {S.~F.}\ \bibnamefont {Yelin}}, \bibinfo {author}
  {\bibfnamefont {M.~D.}\ \bibnamefont {Lukin}}, \ and\ \bibinfo {author}
  {\bibfnamefont {J.~I.}\ \bibnamefont {Cirac}},\ }\href {\doibase
  10.1103/PhysRevA.86.012116} {\bibfield  {journal} {\bibinfo  {journal} {Phys.
  Rev. A}\ }\textbf {\bibinfo {volume} {86}},\ \bibinfo {pages} {012116}
  (\bibinfo {year} {2012})}\BibitemShut {NoStop}%
\bibitem [{\citenamefont {Bhaseen}\ \emph {et~al.}(2012)\citenamefont
  {Bhaseen}, \citenamefont {Mayoh}, \citenamefont {Simons},\ and\ \citenamefont
  {Keeling}}]{Keeling1}%
  \BibitemOpen
  \bibfield  {author} {\bibinfo {author} {\bibfnamefont {M.~J.}\ \bibnamefont
  {Bhaseen}}, \bibinfo {author} {\bibfnamefont {J.}~\bibnamefont {Mayoh}},
  \bibinfo {author} {\bibfnamefont {B.~D.}\ \bibnamefont {Simons}}, \ and\
  \bibinfo {author} {\bibfnamefont {J.}~\bibnamefont {Keeling}},\ }\href
  {\doibase 10.1103/PhysRevA.85.013817} {\bibfield  {journal} {\bibinfo
  {journal} {Phys. Rev. A}\ }\textbf {\bibinfo {volume} {85}},\ \bibinfo
  {pages} {013817} (\bibinfo {year} {2012})}\BibitemShut {NoStop}%
\bibitem [{\citenamefont {Marcuzzi}\ \emph {et~al.}(2014)\citenamefont
  {Marcuzzi}, \citenamefont {Levi}, \citenamefont {Diehl}, \citenamefont
  {Garrahan},\ and\ \citenamefont {Lesanovsky}}]{JuanPDPT}%
  \BibitemOpen
  \bibfield  {author} {\bibinfo {author} {\bibfnamefont {M.}~\bibnamefont
  {Marcuzzi}}, \bibinfo {author} {\bibfnamefont {E.}~\bibnamefont {Levi}},
  \bibinfo {author} {\bibfnamefont {S.}~\bibnamefont {Diehl}}, \bibinfo
  {author} {\bibfnamefont {J.~P.}\ \bibnamefont {Garrahan}}, \ and\ \bibinfo
  {author} {\bibfnamefont {I.}~\bibnamefont {Lesanovsky}},\ }\href {\doibase
  10.1103/PhysRevLett.113.210401} {\bibfield  {journal} {\bibinfo  {journal}
  {Phys. Rev. Lett.}\ }\textbf {\bibinfo {volume} {113}},\ \bibinfo {pages}
  {210401} (\bibinfo {year} {2014})}\BibitemShut {NoStop}%
\bibitem [{\citenamefont {Casteels}\ \emph {et~al.}(2017)\citenamefont
  {Casteels}, \citenamefont {Fazio},\ and\ \citenamefont {Ciuti}}]{Fazio}%
  \BibitemOpen
  \bibfield  {author} {\bibinfo {author} {\bibfnamefont {W.}~\bibnamefont
  {Casteels}}, \bibinfo {author} {\bibfnamefont {R.}~\bibnamefont {Fazio}}, \
  and\ \bibinfo {author} {\bibfnamefont {C.}~\bibnamefont {Ciuti}},\ }\href
  {\doibase 10.1103/PhysRevA.95.012128} {\bibfield  {journal} {\bibinfo
  {journal} {Phys. Rev. A}\ }\textbf {\bibinfo {volume} {95}},\ \bibinfo
  {pages} {012128} (\bibinfo {year} {2017})}\BibitemShut {NoStop}%
\bibitem [{\citenamefont {Collura}\ \emph {et~al.}(2018)\citenamefont
  {Collura}, \citenamefont {De~Luca},\ and\ \citenamefont
  {Viti}}]{collura2018analytic}%
  \BibitemOpen
  \bibfield  {author} {\bibinfo {author} {\bibfnamefont {M.}~\bibnamefont
  {Collura}}, \bibinfo {author} {\bibfnamefont {A.}~\bibnamefont {De~Luca}}, \
  and\ \bibinfo {author} {\bibfnamefont {J.}~\bibnamefont {Viti}},\ }\href@noop
  {} {\bibfield  {journal} {\bibinfo  {journal} {Physical Review B}\ }\textbf
  {\bibinfo {volume} {97}},\ \bibinfo {pages} {081111} (\bibinfo {year}
  {2018})}\BibitemShut {NoStop}%
\bibitem [{\citenamefont {Gamayun}\ \emph {et~al.}(2019)\citenamefont
  {Gamayun}, \citenamefont {Miao},\ and\ \citenamefont
  {Ilievski}}]{gamayun2019domain}%
  \BibitemOpen
  \bibfield  {author} {\bibinfo {author} {\bibfnamefont {O.}~\bibnamefont
  {Gamayun}}, \bibinfo {author} {\bibfnamefont {Y.}~\bibnamefont {Miao}}, \
  and\ \bibinfo {author} {\bibfnamefont {E.}~\bibnamefont {Ilievski}},\
  }\href@noop {} {\bibfield  {journal} {\bibinfo  {journal} {Physical Review
  B}\ }\textbf {\bibinfo {volume} {99}},\ \bibinfo {pages} {140301} (\bibinfo
  {year} {2019})}\BibitemShut {NoStop}%
\bibitem [{\citenamefont {Misguich}\ \emph {et~al.}(2019)\citenamefont
  {Misguich}, \citenamefont {Pavloff},\ and\ \citenamefont
  {Pasquier}}]{misguich2019domain}%
  \BibitemOpen
  \bibfield  {author} {\bibinfo {author} {\bibfnamefont {G.}~\bibnamefont
  {Misguich}}, \bibinfo {author} {\bibfnamefont {N.}~\bibnamefont {Pavloff}}, \
  and\ \bibinfo {author} {\bibfnamefont {V.}~\bibnamefont {Pasquier}},\
  }\href@noop {} {\bibfield  {journal} {\bibinfo  {journal} {SciPost Phys.}\
  }\textbf {\bibinfo {volume} {7}},\ \bibinfo {pages} {025} (\bibinfo {year}
  {2019})}\BibitemShut {NoStop}%
\bibitem [{\citenamefont {Medenjak}\ and\ \citenamefont
  {De~Nardis}(2020)}]{medenjak2020domain}%
  \BibitemOpen
  \bibfield  {author} {\bibinfo {author} {\bibfnamefont {M.}~\bibnamefont
  {Medenjak}}\ and\ \bibinfo {author} {\bibfnamefont {J.}~\bibnamefont
  {De~Nardis}},\ }\href@noop {} {\bibfield  {journal} {\bibinfo  {journal}
  {Physical Review B}\ }\textbf {\bibinfo {volume} {101}},\ \bibinfo {pages}
  {081411} (\bibinfo {year} {2020})}\BibitemShut {NoStop}%
\bibitem [{\citenamefont {Collura}\ \emph {et~al.}(2020)\citenamefont
  {Collura}, \citenamefont {De~Luca}, \citenamefont {Calabrese},\ and\
  \citenamefont {Dubail}}]{collura2020domain}%
  \BibitemOpen
  \bibfield  {author} {\bibinfo {author} {\bibfnamefont {M.}~\bibnamefont
  {Collura}}, \bibinfo {author} {\bibfnamefont {A.}~\bibnamefont {De~Luca}},
  \bibinfo {author} {\bibfnamefont {P.}~\bibnamefont {Calabrese}}, \ and\
  \bibinfo {author} {\bibfnamefont {J.}~\bibnamefont {Dubail}},\ }\href@noop {}
  {\bibfield  {journal} {\bibinfo  {journal} {arXiv preprint arXiv:2001.04948}\
  } (\bibinfo {year} {2020})}\BibitemShut {NoStop}%
\bibitem [{\citenamefont {Breuer}\ \emph {et~al.}(2002)\citenamefont {Breuer},
  \citenamefont {Petruccione} \emph {et~al.}}]{BPTextbook}%
  \BibitemOpen
  \bibfield  {author} {\bibinfo {author} {\bibfnamefont {H.-P.}\ \bibnamefont
  {Breuer}}, \bibinfo {author} {\bibfnamefont {F.}~\bibnamefont {Petruccione}},
   \emph {et~al.},\ }\href@noop {} {\emph {\bibinfo {title} {The theory of open
  quantum systems}}}\ (\bibinfo  {publisher} {Oxford University Press on
  Demand},\ \bibinfo {year} {2002})\BibitemShut {NoStop}%
\bibitem [{\citenamefont {Gardiner}\ \emph {et~al.}(2004)\citenamefont
  {Gardiner}, \citenamefont {Zoller},\ and\ \citenamefont
  {Zoller}}]{GardinerTextbook}%
  \BibitemOpen
  \bibfield  {author} {\bibinfo {author} {\bibfnamefont {C.}~\bibnamefont
  {Gardiner}}, \bibinfo {author} {\bibfnamefont {P.}~\bibnamefont {Zoller}}, \
  and\ \bibinfo {author} {\bibfnamefont {P.}~\bibnamefont {Zoller}},\
  }\href@noop {} {\emph {\bibinfo {title} {Quantum noise: a handbook of
  Markovian and non-Markovian quantum stochastic methods with applications to
  quantum optics}}}\ (\bibinfo  {publisher} {Springer Science \& Business
  Media},\ \bibinfo {year} {2004})\BibitemShut {NoStop}%
\bibitem [{\citenamefont {Torres}(2014)}]{Torres1}%
  \BibitemOpen
  \bibfield  {author} {\bibinfo {author} {\bibfnamefont {J.~M.}\ \bibnamefont
  {Torres}},\ }\href {\doibase 10.1103/PhysRevA.89.052133} {\bibfield
  {journal} {\bibinfo  {journal} {Phys. Rev. A}\ }\textbf {\bibinfo {volume}
  {89}},\ \bibinfo {pages} {052133} (\bibinfo {year} {2014})}\BibitemShut
  {NoStop}%
\bibitem [{\citenamefont {Briegel}\ and\ \citenamefont
  {Englert}(1993)}]{Englert}%
  \BibitemOpen
  \bibfield  {author} {\bibinfo {author} {\bibfnamefont {H.-J.}\ \bibnamefont
  {Briegel}}\ and\ \bibinfo {author} {\bibfnamefont {B.-G.}\ \bibnamefont
  {Englert}},\ }\href {\doibase 10.1103/PhysRevA.47.3311} {\bibfield  {journal}
  {\bibinfo  {journal} {Phys. Rev. A}\ }\textbf {\bibinfo {volume} {47}},\
  \bibinfo {pages} {3311} (\bibinfo {year} {1993})}\BibitemShut {NoStop}%
\bibitem [{SM()}]{SM}%
  \BibitemOpen
  \href@noop {} {}\bibinfo {howpublished} {Supplemental Material}\BibitemShut
  {NoStop}%
\bibitem [{\citenamefont {Bethe}(1931)}]{Bethe1}%
  \BibitemOpen
  \bibfield  {author} {\bibinfo {author} {\bibfnamefont {H.}~\bibnamefont
  {Bethe}},\ }\href {\doibase 10.1007/BF01341708} {\bibfield  {journal}
  {\bibinfo  {journal} {Zeitschrift f{\"u}r Physik}\ }\textbf {\bibinfo
  {volume} {71}},\ \bibinfo {pages} {205} (\bibinfo {year} {1931})}\BibitemShut
  {NoStop}%
\bibitem [{\citenamefont {Korepin}\ \emph {et~al.}(1997)\citenamefont
  {Korepin}, \citenamefont {Bogoliubov},\ and\ \citenamefont
  {Izergin}}]{Bethe2}%
  \BibitemOpen
  \bibfield  {author} {\bibinfo {author} {\bibfnamefont {V.~E.}\ \bibnamefont
  {Korepin}}, \bibinfo {author} {\bibfnamefont {N.~M.}\ \bibnamefont
  {Bogoliubov}}, \ and\ \bibinfo {author} {\bibfnamefont {A.~G.}\ \bibnamefont
  {Izergin}},\ }\href@noop {} {\emph {\bibinfo {title} {Quantum inverse
  scattering method and correlation functions}}},\ Vol.~\bibinfo {volume} {3}\
  (\bibinfo  {publisher} {Cambridge university press},\ \bibinfo {year}
  {1997})\BibitemShut {NoStop}%
\bibitem [{\citenamefont {Parmee}\ and\ \citenamefont {Cooper}(2018)}]{Chris}%
  \BibitemOpen
  \bibfield  {author} {\bibinfo {author} {\bibfnamefont {C.~D.}\ \bibnamefont
  {Parmee}}\ and\ \bibinfo {author} {\bibfnamefont {N.~R.}\ \bibnamefont
  {Cooper}},\ }\href {\doibase 10.1103/PhysRevA.97.053616} {\bibfield
  {journal} {\bibinfo  {journal} {Phys. Rev. A}\ }\textbf {\bibinfo {volume}
  {97}},\ \bibinfo {pages} {053616} (\bibinfo {year} {2018})}\BibitemShut
  {NoStop}%
\bibitem [{\citenamefont {Parmee}\ and\ \citenamefont {Cooper}(2019)}]{Chris2}%
  \BibitemOpen
  \bibfield  {author} {\bibinfo {author} {\bibfnamefont {C.~D.}\ \bibnamefont
  {Parmee}}\ and\ \bibinfo {author} {\bibfnamefont {N.~R.}\ \bibnamefont
  {Cooper}},\ }\href {\doibase 10.1103/PhysRevA.99.063615} {\bibfield
  {journal} {\bibinfo  {journal} {Phys. Rev. A}\ }\textbf {\bibinfo {volume}
  {99}},\ \bibinfo {pages} {063615} (\bibinfo {year} {2019})}\BibitemShut
  {NoStop}%
\bibitem [{\citenamefont {Nakagawa}\ \emph {et~al.}(2020)\citenamefont
  {Nakagawa}, \citenamefont {Kawakami},\ and\ \citenamefont {Ueda}}]{Ueda}%
  \BibitemOpen
  \bibfield  {author} {\bibinfo {author} {\bibfnamefont {M.}~\bibnamefont
  {Nakagawa}}, \bibinfo {author} {\bibfnamefont {N.}~\bibnamefont {Kawakami}},
  \ and\ \bibinfo {author} {\bibfnamefont {M.}~\bibnamefont {Ueda}},\
  }\href@noop {} {\enquote {\bibinfo {title} {Exact liouvillian spectrum of a
  one-dimensional dissipative hubbard model},}\ } (\bibinfo {year} {2020}),\
  \Eprint {http://arxiv.org/abs/2003.14202} {arXiv:2003.14202
  [cond-mat.quant-gas]} \BibitemShut {NoStop}%
\bibitem [{\citenamefont {Prosen}\ and\ \citenamefont
  {{\v{Z}}unkovi{\v{c}}}(2010)}]{Bojan3}%
  \BibitemOpen
  \bibfield  {author} {\bibinfo {author} {\bibfnamefont {T.}~\bibnamefont
  {Prosen}}\ and\ \bibinfo {author} {\bibfnamefont {B.}~\bibnamefont
  {{\v{Z}}unkovi{\v{c}}}},\ }\href {\doibase 10.1088/1367-2630/12/2/025016}
  {\bibfield  {journal} {\bibinfo  {journal} {New Journal of Physics}\ }\textbf
  {\bibinfo {volume} {12}},\ \bibinfo {pages} {025016} (\bibinfo {year}
  {2010})}\BibitemShut {NoStop}%
\bibitem [{\citenamefont {Mendoza-Arenas}\ \emph {et~al.}(2019)\citenamefont
  {Mendoza-Arenas}, \citenamefont {\ifmmode \check{Z}\else
  \v{Z}\fi{}nidari\ifmmode~\check{c}\else \v{c}\fi{}}, \citenamefont {Varma},
  \citenamefont {Goold}, \citenamefont {Clark},\ and\ \citenamefont
  {Scardicchio}}]{Clark1}%
  \BibitemOpen
  \bibfield  {author} {\bibinfo {author} {\bibfnamefont {J.~J.}\ \bibnamefont
  {Mendoza-Arenas}}, \bibinfo {author} {\bibfnamefont {M.}~\bibnamefont
  {\ifmmode \check{Z}\else \v{Z}\fi{}nidari\ifmmode~\check{c}\else
  \v{c}\fi{}}}, \bibinfo {author} {\bibfnamefont {V.~K.}\ \bibnamefont
  {Varma}}, \bibinfo {author} {\bibfnamefont {J.}~\bibnamefont {Goold}},
  \bibinfo {author} {\bibfnamefont {S.~R.}\ \bibnamefont {Clark}}, \ and\
  \bibinfo {author} {\bibfnamefont {A.}~\bibnamefont {Scardicchio}},\ }\href
  {\doibase 10.1103/PhysRevB.99.094435} {\bibfield  {journal} {\bibinfo
  {journal} {Phys. Rev. B}\ }\textbf {\bibinfo {volume} {99}},\ \bibinfo
  {pages} {094435} (\bibinfo {year} {2019})}\BibitemShut {NoStop}%
\bibitem [{\citenamefont {Žnidarič}\ \emph {et~al.}(2017)\citenamefont
  {Žnidarič}, \citenamefont {Mendoza-Arenas}, \citenamefont {Clark},\ and\
  \citenamefont {Goold}}]{Clark2}%
  \BibitemOpen
  \bibfield  {author} {\bibinfo {author} {\bibfnamefont {M.}~\bibnamefont
  {Žnidarič}}, \bibinfo {author} {\bibfnamefont {J.~J.}\ \bibnamefont
  {Mendoza-Arenas}}, \bibinfo {author} {\bibfnamefont {S.~R.}\ \bibnamefont
  {Clark}}, \ and\ \bibinfo {author} {\bibfnamefont {J.}~\bibnamefont
  {Goold}},\ }\href {\doibase 10.1002/andp.201600298} {\bibfield  {journal}
  {\bibinfo  {journal} {Annalen der Physik}\ }\textbf {\bibinfo {volume}
  {529}},\ \bibinfo {pages} {1600298} (\bibinfo {year} {2017})}\BibitemShut
  {NoStop}%
\bibitem [{\citenamefont {Mendoza-Arenas}\ \emph {et~al.}(2015)\citenamefont
  {Mendoza-Arenas}, \citenamefont {Clark},\ and\ \citenamefont
  {Jaksch}}]{Clark3}%
  \BibitemOpen
  \bibfield  {author} {\bibinfo {author} {\bibfnamefont {J.~J.}\ \bibnamefont
  {Mendoza-Arenas}}, \bibinfo {author} {\bibfnamefont {S.~R.}\ \bibnamefont
  {Clark}}, \ and\ \bibinfo {author} {\bibfnamefont {D.}~\bibnamefont
  {Jaksch}},\ }\href {\doibase 10.1103/PhysRevE.91.042129} {\bibfield
  {journal} {\bibinfo  {journal} {Phys. Rev. E}\ }\textbf {\bibinfo {volume}
  {91}},\ \bibinfo {pages} {042129} (\bibinfo {year} {2015})}\BibitemShut
  {NoStop}%
\bibitem [{\citenamefont {Mendoza-Arenas}\ \emph {et~al.}(2013)\citenamefont
  {Mendoza-Arenas}, \citenamefont {Grujic}, \citenamefont {Jaksch},\ and\
  \citenamefont {Clark}}]{Clark4}%
  \BibitemOpen
  \bibfield  {author} {\bibinfo {author} {\bibfnamefont {J.~J.}\ \bibnamefont
  {Mendoza-Arenas}}, \bibinfo {author} {\bibfnamefont {T.}~\bibnamefont
  {Grujic}}, \bibinfo {author} {\bibfnamefont {D.}~\bibnamefont {Jaksch}}, \
  and\ \bibinfo {author} {\bibfnamefont {S.~R.}\ \bibnamefont {Clark}},\ }\href
  {\doibase 10.1103/PhysRevB.87.235130} {\bibfield  {journal} {\bibinfo
  {journal} {Phys. Rev. B}\ }\textbf {\bibinfo {volume} {87}},\ \bibinfo
  {pages} {235130} (\bibinfo {year} {2013})}\BibitemShut {NoStop}%
\bibitem [{\citenamefont {Sklyanin}(1988)}]{Sklyanin}%
  \BibitemOpen
  \bibfield  {author} {\bibinfo {author} {\bibfnamefont {E.~K.}\ \bibnamefont
  {Sklyanin}},\ }\href {\doibase 10.1088/0305-4470/21/10/015} {\bibfield
  {journal} {\bibinfo  {journal} {Journal of Physics A: Mathematical and
  General}\ }\textbf {\bibinfo {volume} {21}},\ \bibinfo {pages} {2375}
  (\bibinfo {year} {1988})}\BibitemShut {NoStop}%
\bibitem [{\citenamefont {Ragoucy}(2012)}]{Ragoucy1}%
  \BibitemOpen
  \bibfield  {author} {\bibinfo {author} {\bibfnamefont {E.}~\bibnamefont
  {Ragoucy}},\ }\href {\doibase 10.1088/1742-6596/343/1/012100} {\bibfield
  {journal} {\bibinfo  {journal} {Journal of Physics: Conference Series}\
  }\textbf {\bibinfo {volume} {343}},\ \bibinfo {pages} {012100} (\bibinfo
  {year} {2012})}\BibitemShut {NoStop}%
\bibitem [{\citenamefont {Cramp{\'{e}}}\ \emph {et~al.}(2010)\citenamefont
  {Cramp{\'{e}}}, \citenamefont {Ragoucy},\ and\ \citenamefont
  {Simon}}]{Ragoucy2}%
  \BibitemOpen
  \bibfield  {author} {\bibinfo {author} {\bibfnamefont {N.}~\bibnamefont
  {Cramp{\'{e}}}}, \bibinfo {author} {\bibfnamefont {E.}~\bibnamefont
  {Ragoucy}}, \ and\ \bibinfo {author} {\bibfnamefont {D.}~\bibnamefont
  {Simon}},\ }\href {\doibase 10.1088/1742-5468/2010/11/p11038} {\bibfield
  {journal} {\bibinfo  {journal} {Journal of Statistical Mechanics: Theory and
  Experiment}\ }\textbf {\bibinfo {volume} {2010}},\ \bibinfo {pages} {P11038}
  (\bibinfo {year} {2010})}\BibitemShut {NoStop}%
\bibitem [{\citenamefont {van Tongeren}(2016)}]{van2016introduction}%
  \BibitemOpen
  \bibfield  {author} {\bibinfo {author} {\bibfnamefont {S.~J.}\ \bibnamefont
  {van Tongeren}},\ }\href@noop {} {\bibfield  {journal} {\bibinfo  {journal}
  {Journal of Physics A: Mathematical and Theoretical}\ }\textbf {\bibinfo
  {volume} {49}},\ \bibinfo {pages} {323005} (\bibinfo {year}
  {2016})}\BibitemShut {NoStop}%
\bibitem [{\citenamefont {Yang}\ and\ \citenamefont {Yang}(1969)}]{YangYang}%
  \BibitemOpen
  \bibfield  {author} {\bibinfo {author} {\bibfnamefont {C.-N.}\ \bibnamefont
  {Yang}}\ and\ \bibinfo {author} {\bibfnamefont {C.~P.}\ \bibnamefont
  {Yang}},\ }\href@noop {} {\bibfield  {journal} {\bibinfo  {journal} {Journal
  of Mathematical Physics}\ }\textbf {\bibinfo {volume} {10}},\ \bibinfo
  {pages} {1115} (\bibinfo {year} {1969})}\BibitemShut {NoStop}%
\bibitem [{\citenamefont {Grijalva}\ \emph {et~al.}(2019)\citenamefont
  {Grijalva}, \citenamefont {Nardis},\ and\ \citenamefont {Terras}}]{Jacopo2}%
  \BibitemOpen
  \bibfield  {author} {\bibinfo {author} {\bibfnamefont {S.}~\bibnamefont
  {Grijalva}}, \bibinfo {author} {\bibfnamefont {J.~D.}\ \bibnamefont
  {Nardis}}, \ and\ \bibinfo {author} {\bibfnamefont {V.}~\bibnamefont
  {Terras}},\ }\href {\doibase 10.21468/SciPostPhys.7.2.023} {\bibfield
  {journal} {\bibinfo  {journal} {SciPost Phys.}\ }\textbf {\bibinfo {volume}
  {7}},\ \bibinfo {pages} {23} (\bibinfo {year} {2019})}\BibitemShut {NoStop}%
\bibitem [{\citenamefont {Fendley}(2016)}]{Fendley1}%
  \BibitemOpen
  \bibfield  {author} {\bibinfo {author} {\bibfnamefont {P.}~\bibnamefont
  {Fendley}},\ }\href {\doibase 10.1088/1751-8113/49/30/30lt01} {\bibfield
  {journal} {\bibinfo  {journal} {Journal of Physics A: Mathematical and
  Theoretical}\ }\textbf {\bibinfo {volume} {49}},\ \bibinfo {pages} {30LT01}
  (\bibinfo {year} {2016})}\BibitemShut {NoStop}%
\bibitem [{\citenamefont {Beisert}\ \emph {et~al.}(2013)\citenamefont
  {Beisert}, \citenamefont {Fi{\'{e}}vet}, \citenamefont {de~Leeuw},\ and\
  \citenamefont {Loebbert}}]{nonHermXXZ}%
  \BibitemOpen
  \bibfield  {author} {\bibinfo {author} {\bibfnamefont {N.}~\bibnamefont
  {Beisert}}, \bibinfo {author} {\bibfnamefont {L.}~\bibnamefont
  {Fi{\'{e}}vet}}, \bibinfo {author} {\bibfnamefont {M.}~\bibnamefont
  {de~Leeuw}}, \ and\ \bibinfo {author} {\bibfnamefont {F.}~\bibnamefont
  {Loebbert}},\ }\href {\doibase 10.1088/1742-5468/2013/09/p09028} {\bibfield
  {journal} {\bibinfo  {journal} {Journal of Statistical Mechanics: Theory and
  Experiment}\ }\textbf {\bibinfo {volume} {2013}},\ \bibinfo {pages} {P09028}
  (\bibinfo {year} {2013})}\BibitemShut {NoStop}%
\end{thebibliography}%

\onecolumngrid
\newpage

\renewcommand\thesection{S\arabic{section}}
\renewcommand\theequation{S\arabic{equation}}
\renewcommand\thefigure{S\arabic{figure}}
\setcounter{equation}{0}
\setcounter{figure}{0}

\begin{center}
	{\Large \emph{Supplementary Information}: Exact Solutions of Quantum Many-Body Dynamics Under Loss using the Bethe Ansatz}
\end{center}

\renewcommand\thesection{S\arabic{section}}
\renewcommand\theequation{S\arabic{equation}}
\renewcommand\thefigure{S\arabic{figure}}
\setcounter{equation}{0}
\setcounter{section}{0}
In the supplemental material we provide the discussion of technical details that were omitted from the main text. In the first part we discuss further examples of systems solvable by our method. In the second section we construct the eigenstates of the Liouvillian. In the third section we provide details on the calculation of the Liouvillian gap, and in the final section the results related to the boundary magnons. We also plot the full Liouvillian spectrum for different values of anisotropy $\Delta$ which shows the formation of intriguing band structure (see Fig.~\ref{fig:fullspect}).

\begin{figure*}[h]
    \centering
    \begin{subfigure}{0.4\columnwidth}
        \centering
        \includegraphics[width = 1\columnwidth]{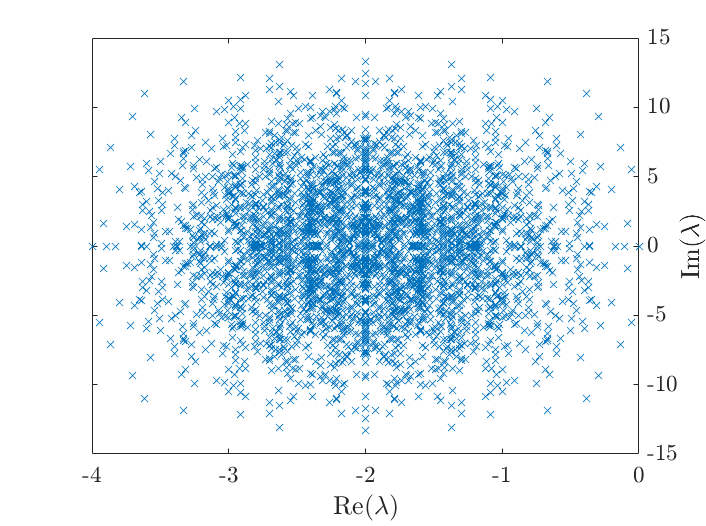}
        \caption{$\Delta = 0.5$}
        \label{fig:fullspect_0.5}
    \end{subfigure}
    \begin{subfigure}{0.4\columnwidth}
    \centering
        \includegraphics[width = 1\columnwidth]{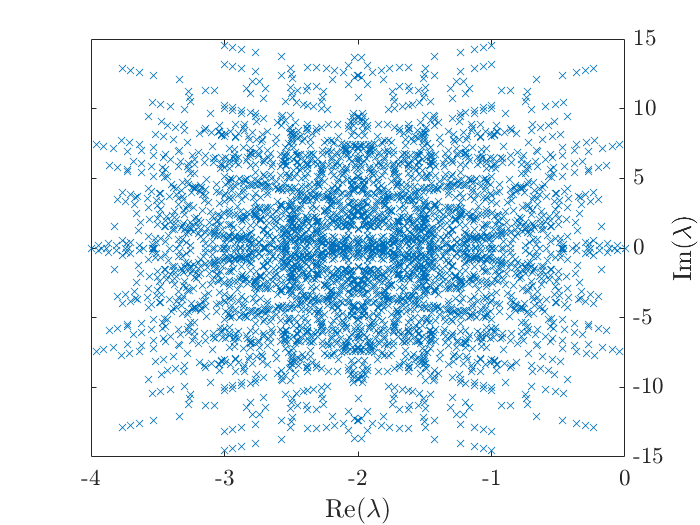}
        \caption{$\Delta = 1.0$}
        \label{fig:fullspect_1.0}
    \end{subfigure}
    \begin{subfigure}{0.4\columnwidth}
    \centering
        \includegraphics[width = 1\columnwidth]{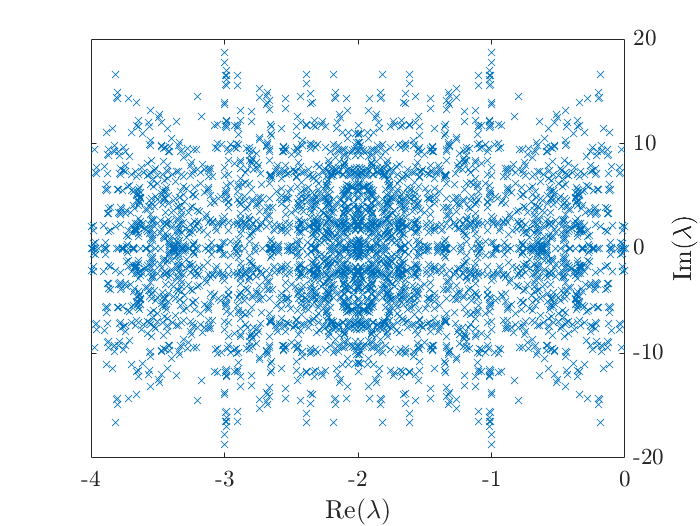}
        \caption{$\Delta = 1.5$}
        \label{fig:fullspect_1.5}
    \end{subfigure}
    \begin{subfigure}{0.4\columnwidth}
    \centering
        \includegraphics[width = 1\columnwidth]{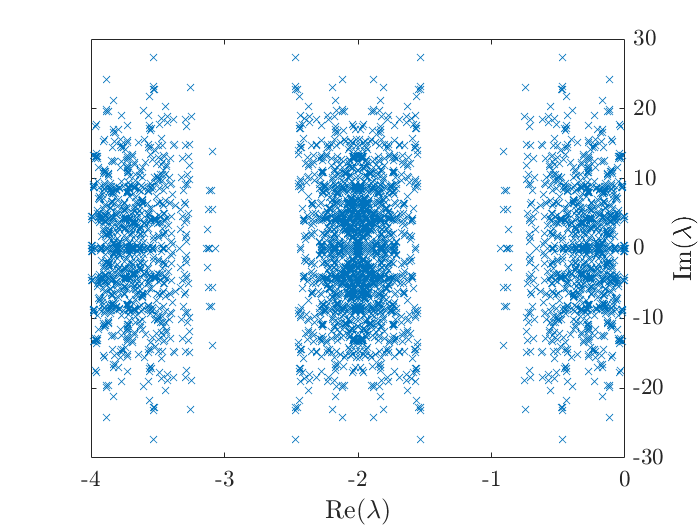}
        \caption{$\Delta = 2.5$}
        \label{fig:fullspect_2.5}
    \end{subfigure}
    \caption{Full spectrum for a 6 site system with $\Gamma = 0.5$ at various $\Delta$. We see the emergence of band structure forming as $\Delta$ increased. This corresponds to different subspaces of the Hilbert space decaying in separate stages.}
    \label{fig:fullspect}
\end{figure*}

\section*{Examples of dissipative quantum models solvable by Bethe ansatz}
Here we provide some further examples of models solvable by the method introduced in the main text.

Consider a general 1D Hamiltonian with raising (lowering) operators for possibly several species, $a_j^\dagger$, $b_j^\dagger,\ldots$ ($a_j,b_j,\ldots$), acting on site $j=1,\ldots, N$. We further assume that $[H,\sum a^\dagger_j a_j]=[H,\sum b^\dagger_j b_j]=0$. We define the dissipative contribution to $\tilde{H}$ as $D\equiv\sum_\mu L_\mu^\dagger L_\mu$. Taking an integrable $H$ we observe that the following types of loss processes render $\tilde{H}$ integrable, 
\begin{enumerate}
\item \label{item1} $L_j=\gamma a_j b_j, j=1,\ldots,N$ (homogenous two-body loss),
\item \label{item3} $L_j=\gamma a^\dagger_j a_{j+1}, L_{j+N-1}=\gamma'a_j a^\dagger_{j+1},  j=1,\ldots,N-1$ (dissipative nearest-neighbor hopping) and,
\item \label{item2} $D=\gamma \sum_{j=1}^{N-1} a^\dagger_{j}a_{j+1} $ (correlated non-local dissipative loss).
\end{enumerate}
The $\tilde{H}$ in cases~\ref{item1}. (\ref{item3}.) are integrable because $D$ may be rewritten as an imaginary interaction term of the form $D=\sum_j a^\dagger_ja_jb^\dagger_jb_j$ ($D=\sum_j\gamma a_j a^\dagger_j a^\dagger_{j+1} a_{j+1}+\gamma' a^\dagger_j a_j a_{j+1} a^\dagger_{j+1}$). In particular, homogeneous two-body loss is a standard loss process in cold atom simulations \cite{coldatomreview,coldexp}. A concrete physical example of this is the 1D Hubbard model with two-body recombination of fermions of spin-down and spin-up \cite{Ueda}. 

The case ~\ref{item2}. has an integrable $\tilde{H}$ because $D$ is just an imaginary contribution to the hopping (kinetic) term in $H$. For instance, for XXZ spin chains such a non-Hermitian Hamiltonian was solved in \cite{nonHermXXZ}. These cases are realized, for instance, by two-level systems coupled by dipolar interactions and subject to nonlocal dissipation, i.e. decay through optical emissions \cite{Chris,Chris2}. 

\section*{Eigenstates of the boundary loss XXZ Liouvillian}
In this section we will construct the right eigenstates of the Liouvillian
\begin{dmath} \label{eq:doublespaceliouvillianapp}
	\LL=-i(H\otimes\one-\one\otimes H^T) +\sum_k \left(2 L_\mu\otimes L_\mu^*-L_\mu^\dagger L_\mu\otimes \one-\one \otimes (L_\mu^\dagger L_\mu)^T\right).
\end{dmath}
In the case of the XXZ chain with a single loss at the first site. First of all, we will remind the reader of the basic structure of Bethe eigenstates, which will then serve to construct the eigenstates of the Liouvillian. The eigenstate of the Hamiltonian
\begin{equation}
    \tilde{H}= - \ii H_{XXZ}-\Gamma \sigma^+_1\sigma^-_1 ,
\end{equation}
pertaining to the energy $E_{\phi_\eta^a}$ reads 
\begin{equation}
    \ket{\phi^a_\eta} = \sum_{1\le x_1 <...<x_a\le n}f_{a,\eta}(x_1,..,x_a)\ket{x_1,...,x_a} . 
\end{equation}
Here the set of $\{x_j\}$ indicate the positions of spin-up excitations while the label $a$ corresponds to a given total magnetisation. Finally $\eta$ labels the state within this sector. In terms of Bethe roots $\{k_j\}$, the wave function reads
\begin{gather}
    f(x_1,...,x_m) = \sum_P \varepsilon_P A(k_1,...,k_m)e^{i(k_1x_1+ \dots k_mx_m)} \\
    A(k_1,...,k_m) = \prod_{j=1}^{m} (\Delta e^{-ik_j N} - e^{-i(N+1)k_j}) \prod_{1\le j < l \le m} B(-k_j, k_l)e^{-ik_l} \\
    B(k,k') = (1-2\Delta e^{ik'}+e^{i(k+k')})(1-2\Delta e^{-ik}+e^{i(k'-k)}),
\end{gather}
where the summation is performed over all permutations and negations of $\{k_j\}$, and $\varepsilon_P$ changes sign with each such \emph{mutation}.

We then use the triangular form for the Liouvillian, i.e that
\begin{equation}
    {\sL}: \sM^a \otimes \sM^b \rightarrow \left(\sM^a \otimes \sM^b\right) \oplus \left(\sM^{a-1} \otimes \sM^{b-1}\right) , 
\end{equation}
where $\sM^a$ is the subspace with magnetization $a$, to make the ansatz that the eigenstate of ${\sL}$ with eigenvalue $E_{\phi_\eta^a}+E_{\phi_\zeta^b}^*$ is given by
\begin{equation}
    \ket{\Phi_{\eta,\zeta}^{a,b}} = \ket{\phi_\eta^a}\overline{\ket{\phi_\zeta^b}} + \sum_{\mu = 1}^{\min\{a,b\}} \sum_{i,j} B_\mu(i,j) \ket{\phi_i^{a-\mu}} \overline{\ket{\phi_j^{b-\mu}}} . 
\end{equation}
Substituting this gives the recurrence relation 
\begin{equation}
    B_\mu(i,j) = \frac{8\Gamma}{E_\mu(i,j)} \sum_{p,q} B_{\mu-1}(p,q) \Sigma_{a-\mu+1}(p,i) \Sigma_{b-\mu+1}(q,j)^* \ , \ B_0(i,j) = \delta_{i,\eta}\delta_{j,\zeta}
\end{equation}
for $\mu = 1,...,\min\{a,b\}$, where we defined 
\begin{gather}
    E_\mu(i,j) = (E_{\phi_\eta^a}+E_{\phi_\zeta^b}^*)-(E_{\phi_i^{a-\mu}}+E_{\phi_j^{b-\mu}}^*) \\
    \Sigma_m(p,i) = \sum_{\{x_2,\dots x_m\},\ x_2>1} f_{m,p}(1,x_2,\dots x_m)f_{m-1,i}(x_2,\dots,x_m).
\end{gather}

In general, this recursion relation is significantly more complex than computing the Bethe states of $\tilde{H}$. It does however provide an insight into the structure of the eigenstates. We also note that one of the main powers of Bethe ansatz lies in the thermodynamics and that efficient calculations might still be possible in such a limit.

\section*{Thermodynamic limit of the leading decay rates}
Here we will study the solutions of the Bethe equations that have purely real momenta in the thermodynamic limit $\lim_{N \to \infty} k_j=2 n \pi$. These solutions can be physically understood as free magnons that live in the bulk of the system and only experience the effects of other magnons and the boundary in sub-leading order $1/N$. 

To do this we start with the logarithmic form of the Bethe equations,
\begin{equation}
2 \ii k_j=\frac{1 }{N}\left(\sum_{i \neq j} \log\left[S(e^{\ii k_j},e^{\ii k_i})\right] \right)-\frac{2 \ii \pi  I_j}{N}. \label{LogBESUP}
\end{equation}
In order to simplify discussion we focus on the single magnon case, though the solutions for $m$ magnons are also straightforward in the above discussed limit,
\begin{equation}
2 \ii k=\frac{\Omega(e^{\ii k}) }{N}-\frac{2 \ii \pi  I_1}{N}, \label{singmangSUP}  
\end{equation}

We denote by 
\begin{equation}
    \Omega(a)= \log \left(\frac{(a \Delta -1) (-1+a (\Delta -2 \ii \Gamma ))}{(a-\Delta ) (a+2 \ii \Gamma -\Delta )}\right),
\end{equation}
and expand the momenta $k$ as the power series in $1/N$, $k=k^{(0)}+1/N k^{(1)}+\ldots$, which we truncate at the order $\mathcal{O}(1/N^3)$. It is important to distinguish the cases when the integer $I_1$ is finite and when it is of the order of the system size $N$ or close to  $N$. Since the leading decay mode corresponds to the latter case, we make the transformation $I_1 \to N-I_1$ and focus on finite $I_1$. 

Expanding \eqref{singmangSUP} is straightforward, as is solving it order by order. We arrive at,
\begin{dmath}
k=-\pi+\frac{1}{N}\pi  I_1+\frac{1}{N^2}\pi  I_1 \left(-\frac{1}{-2 i \Gamma +\Delta +1}-\frac{1}{\Delta
  +1}+1\right)+\frac{1}{N^3}\frac{\pi  I_1 \left(-2 i \Gamma  \Delta +\Delta ^2-1\right)^2}{(\Delta +1)^2  (-2 i \Gamma +\Delta +1)^2}+{\cal{O}}\left(\frac{1}{N^4}\right). \label{SUPPsolutionsK}
\end{dmath}
Let us recall the eigenvalue equation,
\begin{equation} 
    \lambda_{i,j} = -4\ii (\cos(k_i)-\cos(k_j^*)),
\end{equation}
where $i,j$ distinguishes different solutions given by $I_1$ in \eqref{SUPPsolutionsK}. The gap comes from an off-diagonal state composed of the vacuum state and the single spin-up excitation, i.e. $k_i=0, k_j=k(I_1=1)$. Setting this gives the gap equation in the main text. 

We will now study the one top-magnon (spin-down in a background of all spins up) sector.

\subsection*{Calculation of the phase transition in the highly excited eigenstates}

The single top-magnon cases correspond to setting $\Omega(a)= \log \left(\frac{(a \Delta -1) (-1+a (\Delta +2 \ii \Gamma ))}{(a-\Delta ) (a-2 \ii \Gamma -\Delta )}\right)$ in \eqref{singmangSUP}. The corresponding energies of $\tilde{H}$ are,
\begin{equation}
E_p=-4 \Gamma -4 \ii (\cos(k_p)-\Delta), \label{energySUP}
\end{equation}
while the eigenvalues of $\LL$ read,
\begin{equation}
\lambda_{p,p^'}=E_p+E^*_{p'}. \label{eigenvalueSUP}
\end{equation}
We now look at the case when $I_1$ is finite. Numerically we observe that this corresponds to the leading decay rates in the single top-magnon sector for $\Delta<1$ in the small $\Gamma$ limit. Performing the same expansion as before, relabeling $I_1 \rightarrow p$, we obtain (now it is sufficient to look at only up to order $1/N^2$),
\begin{equation}
k_{p}=-\frac{p \pi}{N}+\frac{1}{N^2} \left[\pi  I_1 \left(-\frac{1}{-2 \ii \Gamma +\Delta -1}+\frac{1}{1-\Delta }-1\right)\right]. \label{expansionSUP}
\end{equation}
We take the derivative w.r.t to $\Delta$ of \eqref{eigenvalueSUP},
\begin{equation}
\frac{d \lambda_{p,p'}}{d \Delta}=-4 \ii[-\sin(k_p) \frac{dk_p}{d\Delta}+\sin(k_{p'}^*) \frac{dk_{p'}^*}{d\Delta}].
\end{equation}
Using \eqref{expansionSUP} we obtain,
\begin{equation}
    \frac{d \lambda_{p,p'}}{d \Delta}=-4 i\pi^2 \left(\frac{p^2-{p'}^2}{(1-\Delta )^2}+\frac{p^2+{p'}^2}{(-2 i \Gamma +\Delta -1)^2}\right)  \frac{1}{N^3} + O\left(\frac{1}{N^4}\right)
\end{equation}
which diverges as $\Delta \to 1$ at leading order in $\Gamma$, signaling the phase transition in these highly excited states. 

\section*{Boundary bound modes and stability of the domain wall in the easy-axis regime}

In the easy-axis regime, $\Delta>1$, an infinite number of solutions that have non-zero imaginary part in the thermodynamic limit appear, $\lim_{N \to \infty}\Im(k_j) \neq 0$. In contrast, in the easy-plane regime we observe that only a finite number of such solutions can appear at finite $\Gamma \ne 0$. 

We study the solutions with $\lim_{N \to \infty}\Im(k_j)>0$. These correspond to top-magnons localized at the boundary loss site, which we refer to as boundary bound modes. More specifically, we may solve the $m$ top-magnon Bethe equations in the $N \to \infty$ limit by observing that $e^{\ii k_j N} \to 0$. Focusing on the top-magnon boundary bound modes, we arrive at the following simple form of the Bethe equations in the $N \to \infty$ limit,
\begin{equation} \label{BEQSUP}
\left(1+e^{\ii k_j} (-\Delta -2 \ii \Gamma )\right) \prod_{i
\neq j}\left(-2 \Delta  e^{\ii k_j}+e^{\ii (k_j-k_i)}+1\right) \left(e^{\ii (k_j+k_i)}-2 \Delta  e^{\ii
  k_j}+1\right) =0.
\end{equation}
These may be recursively solved,
\begin{equation}
\label{relationsSUP}
\exp(-\ii k_j)+\exp(i k_{j-1})=2 \Delta,\quad\exp(-\ii k_1)= \Delta+2\ii \Gamma.
\end{equation}
Physically, this means that the $k_1$ top-magnon is localized at the loss site, whereas the $j$-th top-magnon is bound to the $(j-1)$-st one. This characterises a domain wall state. 

We will now show that the real part of the eigenvalues decay exponentially with top-magnon number $m$. 
First, we decompose the equations for energies \eqref{energySUP} as
\begin{equation}
E_m=-4 \Gamma-4\ii \sum_{i=1}^m\left(\frac{1}{2}\left(\exp(\ii k_i)+\exp(-\ii k_i)\right)-\Delta\right).
\end{equation}
Regrouping the terms, we simply get
\begin{equation}
E_m=-2\ii \left(\exp(\ii k_m)-\Delta\right).
\end{equation}
In order to demonstrate stability it is sufficient to show that the imaginary part of $ \exp(\ii k_m) $ goes to $ 0 $ in the limit of large top-magnon number, $ m\to \infty $. Let us write recursion relations for  $\exp(\ii k_m) $
\begin{equation}
\exp(\ii k_{2j+1})=-\frac{2\Delta-\exp(\ii k_{2j-1})}{1-2\Delta(2\Delta- \exp(\ii k_{2j-1}))}, \quad \exp(\ii k_1)= (\Delta+2\ii \Gamma)^{-1}, \quad \exp(ik_2) = \frac{\Delta - 2i \Gamma}{2\Delta(\Delta-2i \Gamma)-1}.
\end{equation}
Solving for the stationary value of recursion, $ z=\exp(\ii k_{2j+1})=\exp(\ii k_{2j-1}) $, we obtain two \emph{real} solutions,
\begin{equation}
z_1=\Delta-\sqrt{-1+\Delta^2},\quad z_2=\Delta+\sqrt{-1+\Delta^2},
\end{equation}
with the stable point being $ z_1 $.
We numerically observe that this fixed point is converged to for any initial value of $\Delta>1$ and $\Gamma$. 
The decay of the most stable eigenvalue $\lambda_m=2\Re(E_m)$ is thus exponential in top-magnon number $m$, demonstrating the stability of the domain wall.

\end{document}